\documentclass[twocolumn]{aastex631}
\usepackage{upgreek}
\usepackage{amsmath}
\usepackage[shortcuts]{extdash}

\newcommand{\jladd}[1]{{#1}}
\usepackage{csquotes}
\SetBlockThreshold{0}

\shorttitle{Improved companion mass limits for Sirius A with thermal infrared coronagraphy}
\shortauthors{Long et al.}
\graphicspath{{./}{./}}

\newcommand{\um}{\ensuremath{\upmu}\textrm{m}}
\newcommand{\Rsun}{\ensuremath{R_\odot}}
\newcommand{\Lsun}{\ensuremath{L_\odot}}
\newcommand{\omigru}{o Gruis}
\begin{document}

\title{Improved companion mass limits for Sirius A with thermal infrared coronagraphy using a vector-apodizing phase plate and time-domain starlight subtraction techniques}
\correspondingauthor{Joseph D. Long}
\email{josephlong@arizona.edu}

\author[0000-0003-1905-9443]{Joseph D. Long}
\affiliation{Steward Observatory, The University of Arizona, Tucson, AZ, USA}

\author[0000-0002-2346-3441]{Jared R. Males}
\affiliation{Steward Observatory, The University of Arizona, Tucson, AZ, USA}

\author[0000-0001-5130-9153]{Sebastiaan Y. Haffert}
\affiliation{Steward Observatory, The University of Arizona, Tucson, AZ, USA}

\author[0000-0003-3904-7378]{Logan Pearce}
\affiliation{Steward Observatory, The University of Arizona, Tucson, AZ, USA}

\author[0000-0002-5251-2943]{Mark S. Marley}
\affiliation{Lunar \& Planetary Laboratory, The University of Arizona, Tucson, AZ, USA}

\author[0000-0002-1384-0063]{Katie M. Morzinski}
\affiliation{Steward Observatory, The University of Arizona, Tucson, AZ, USA}

\author[0000-0002-2167-8246]{Laird M. Close}
\affiliation{Steward Observatory, The University of Arizona, Tucson, AZ, USA}

\author[0000-0002-6717-1977]{Gilles P. P. L. Otten}
\affiliation{Leiden Observatory, Leiden University, Leiden, The Netherlands}
\affiliation{Academia Sinica, Institute of Astronomy and Astrophysics, Taipei, Taiwan}

\author[0000-0003-1946-7009]{Frans Snik}
\affiliation{Leiden Observatory, Leiden University, Leiden, The Netherlands}

\author[0000-0002-7064-8270]{Matthew A. Kenworthy}
\affiliation{Leiden Observatory, Leiden University, Leiden, The Netherlands}

\author[0000-0002-1368-841X]{Christoph U. Keller}
\affiliation{Leiden Observatory, Leiden University, Leiden, The Netherlands}
\affiliation{Lowell Observatory, Flagstaff, AZ, USA}

\author[0000-0002-1954-4564]{Philip Hinz}
\affiliation{Steward Observatory, University of Arizona, Tucson, AZ, USA}
\affiliation{University of California Observatories, Santa Cruz, CA, USA}

\author[0000-0002-3380-3307]{John D. Monnier}
\affiliation{University of Michigan, Department of Astronomy, Ann Arbor, MI, USA}

\author[0000-0001-6654-7859]{Alycia Weinberger}
\affiliation{Carnegie Institution of Washington, Earth and Planets Laboratory, Washington D.C., USA}

\author[0000-0003-1841-2241]{Volker Tolls}
\affiliation{Harvard-Smithsonian Center for Astrophysics, Cambridge, MA, USA}

\begin{abstract}
%The search for companions to Sirius has a long history, beginning with the prediction of a companion from proper motion measurements in 1844 and observational confirmation of the existence of Sirius B eighteen years later. Modern adaptive optics systems have been applied to a succession of high-contrast imaging observations searching for additional brown dwarf and planetary companions.
We use observations with the infrared-optimized MagAO system and Clio camera in 3.9 \um{} light to place stringent mass constraints on possible undetected companions to Sirius A. We suppress the light from Sirius A by imaging it through a grating vector-apodizing phase plate coronagraph with 180º dark region (gvAPP-180). To remove residual starlight in post-processing, we apply a time-domain principal-components-analysis-based algorithm we call PCA-Temporal (PCAT), which uses eigen-time-series rather than eigen-images to subtract starlight. By casting the problem in terms of eigen-time-series, we reduce the computational cost of post-processing the data, enabling the use of the fully sampled dataset for improved contrast at small separations.  We also discuss the impact of retaining fine temporal sampling of the data on final contrast limits. We achieve post-processed contrast limits of $1.5 \times 10^{-6}$ to $9.8 \times 10^{-6}$ outside of 0.75 arcsec which correspond to planet masses of 2.6 to 8.0 $M_J$. These are combined with values from the recent literature of high-contrast imaging observations of Sirius to synthesize an overall completeness fraction as a function of mass and separation. After synthesizing these recent studies and our results, the final completeness analysis rules out 99\% of $\ge 9 \ M_J$ planets from 2.5--7 AU.
\end{abstract}


\section{Introduction} \label{sec:intro}

As the brightest star in the sky, it is no surprise that Sirius has been observed since antiquity. Measurements of the proper motion of Sirius in the 1800s led to the prediction that a second body was affecting the motion of the Sirius system \citep{Bessel1844}.
Eighteen years later, this was confirmed observationally with the detection of Sirius B \citep{Bond1862}.
When a 6.25 year periodic perturbation was noted and analyzed by \cite{Benest1995}, there was a renewed search for a low-mass companion to Sirius A that could match their prediction of a mass $M \lesssim 50 M_J$ object orbiting at $\sim 7.9$ AU.
Studies have been undertaken with HST \citep{Schroeder2000} and from the ground with adaptive optics systems (\citealt{Thalmann2011}, \citealt{Vigan2015}, among others) and most recently by the VISIR team in 10 \um{} light \citep{pathakHighcontrastImagingTen2021}.
These observations have ruled out the existence of low-mass companions in the mass-separation regime predicted by Benest \& Duvent, but Sirius remains an attractive target for exoplanet direct imaging searches for other reasons.

The proximity of Sirius \citep[d = 2.67 pc,][]{gaiaDistances} means that smaller projected orbital radii are accessible to direct imaging instruments than on more distant stars.
Furthermore, extrasolar giant planets are statistically not uncommon around ``retired A stars'' \citep{Johnson2007,Ghezzi2018}, \jladd{and we expect this to also be true for A-type progenitors like Sirius.} Probing smaller separations and longer wavelength ranges increases the chance of finding a low-mass companion previous studies may have missed.
The high $T_\mathrm{eff} = 9910 \pm 130$ K \citep{liebertAgeProgenitorMass2005} of Sirius A means the contribution of \jladd{instellation} to the effective temperatures of planets at small separations will be significant, boosting their near-infrared luminosity and thus our ability to detect them. We summarize the properties of Sirius A and the relevant sources in Table~\ref{tab:sirius}.

\begin{deluxetable*}{rccc}
\tablenum{1}
\tablecaption{Properties of Sirius A\label{tab:sirius}}
\tablewidth{0pt}
\tablehead{
\colhead{Property} & \colhead{Symbol} & \colhead{Value} & \colhead{Source}
}
\startdata
age                   & $t$ & 242 $\pm$ 15 Myr & \cite{Bond2017} \\
luminosity            & $L_\mathrm{Sirius}$ & $25.4 \pm 1.3$ \Lsun & \cite{liebertAgeProgenitorMass2005} \\
radius                & $R_\mathrm{Sirius} $ & $1.711 \pm 0.013$ \Rsun & \cite{kervellaSirius} \\
effective temperature & $T_\mathrm{Sirius}$ & 9910 $\pm$ 130 K & from $L_\mathrm{Sirius}$ and $R_\mathrm{Sirius} $ \\
distance from Earth   & $d_\mathrm{Sirius} $ & 2.67 $\pm$ 0.001 pc & \cite{gaiaDistances} \\
magnitude in Clio [3.95] filter & $m_\mathrm{[3.95]}$  & -1.39 & using \cite{calspecSiriusUpdate2019} spectrum \\
\enddata
% \tablecomments{}
\end{deluxetable*}

\section{Observations} \label{sec:observations}

\begin{deluxetable*}{rccccc}
    \tablenum{2}
    \tablecaption{Observations of Sirius A and \omigru{}: Conditions\label{tab:conditions}}
    \tablewidth{0pt}
    \tablehead{
    \colhead{Target} & \colhead{Start UT} & \colhead{End UT} & \colhead{Sky Rotation} & \colhead{Seeing [$''$]} & \colhead{PWV [mm]}
    }
    \startdata
    \omigru{} & 2015-11-29 00:38:01 & 2015-11-29 00:45:10 & --- & 0.58 & 2.0\\
    Sirius A & 2015-11-29 05:58:17 & 2015-11-29 08:55:32 & $105^\circ$ & 0.39--0.53 & 1.9-2.2\\
    % 2015-11-29 & \omigru{} & [3.95] & 0.5 &  & \\
    % age                   & $t$ & 242 $\pm$ 15 Myr & \cite{Bond2017} \\
    % luminosity            & $L_\mathrm{Sirius}$ & $25.4 \pm 1.3$ \Lsun & \cite{liebertAgeProgenitorMass2005} \\
    % radius                & $R_\mathrm{Sirius} $ & $1.711 \pm 0.013$ \Rsun & \cite{kervellaSirius} \\
    % effective temperature & $T_\mathrm{Sirius}$ & 9910 $\pm$ 130 K & from $L_\mathrm{Sirius}$ and $R_\mathrm{Sirius} $ \\
    % distance from Earth   & $d_\mathrm{Sirius} $ & 2.67 $\pm$ 0.001 pc & \cite{gaiaDistances} \\
    % magnitude in Clio [3.95] filter & $m_\mathrm{[3.95]}$  & -1.39 & using \cite{calspecSiriusUpdate2019} spectrum \\
    \enddata
    \tablecomments{PWV measurements are approximate.}
\end{deluxetable*}

\begin{deluxetable*}{rccccccc}
    \tablenum{3}
    \tablecaption{Observations of Sirius A and \omigru{}: Camera \& MagAO configuration\label{tab:observations}}
    \tablewidth{0pt}
    \tablehead{
    \colhead{Target} &\colhead{AO Modes} & \colhead{Loop Freq. [Hz]} & \colhead{Filter} & \colhead{Coronagraph} & \colhead{Purpose} & \colhead{Exposure [s]} & \colhead{Frames}
    }
    \startdata
    \omigru{} & 300 & 989.6 & [3.95] & gvAPP-180 & calibration & 5 & 30 \\
    & & & & & sky & 5 & 30\\
    Sirius A & 300 & 989.6 & [3.95] & gvAPP-180 & science & 0.5 & 10,889 \\
    & & & & & sky & 0.5 & 1220\\
    & & & & & PSF cal.* & 0.28 & 8\\
    & & & & & PSF cal. sky & 0.28 & 8\\
    % \omigru{} & 2015-11-29 00:38:01 & 2015-11-29 00:45:10 & $''$ & 5 & 30 & $''$\\
    % & & & $''$ & $''$ & $''$ & $''$\\
    % 2015-11-29 & \omigru{} & [3.95] & 0.5 &  & \\
    % age                   & $t$ & 242 $\pm$ 15 Myr & \cite{Bond2017} \\
    % luminosity            & $L_\mathrm{Sirius}$ & $25.4 \pm 1.3$ \Lsun & \cite{liebertAgeProgenitorMass2005} \\
    % radius                & $R_\mathrm{Sirius} $ & $1.711 \pm 0.013$ \Rsun & \cite{kervellaSirius} \\
    % effective temperature & $T_\mathrm{Sirius}$ & 9910 $\pm$ 130 K & from $L_\mathrm{Sirius}$ and $R_\mathrm{Sirius} $ \\
    % distance from Earth   & $d_\mathrm{Sirius} $ & 2.67 $\pm$ 0.001 pc & \cite{gaiaDistances} \\
    % magnitude in Clio [3.95] filter & $m_\mathrm{[3.95]}$  & -1.39 & using \cite{calspecSiriusUpdate2019} spectrum \\
    \enddata
    \tablecomments{Frames with stray light artifacts impacting the dark hole regions were excluded from the original 12,800 to obtain 10,889 science frames. *Not used due to saturation.}
\end{deluxetable*}

We observed Sirius with the Magellan Clay telescope at Las Campanas Observatory on the night of 2015\=/11\=/29 UT using the Magellan Adaptive Optics (MagAO) system \citep{closeDIFFRACTIONLIMITEDVISIBLELIGHT2013} and the Clio infrared (IR) imaging instrument \citep{Morzinski2015}.

The grating vector-apodizing phase plate 180$^\circ$ (gvAPP-180) coronagraph design we employed is a further refinement of the original apodizing phase plate of \cite{Kenworthy2007}. By using a liquid crystal material imprinted with the apodizing phase pattern, the phenomenon of vector (or geometric) phase provides a larger range of wavelengths over which the apodization is effective \citep{snik2012vector}. The grating separates oppositely circularly polarized light on the detector, producing two copies of the point-spread function (PSF) with complementary dark holes---and a nearly 360$^\circ$ accessible dark hole region when the two are recombined. The APP family of coronagraphs are pupil-plane optics, meaning alignment and pointing accuracy tolerances are more forgiving than a small-inner-working-angle (IWA) focal plane coronagraph \citep{doelmanVectorapodizingPhasePlate2021}. The gvAPP-180 in MagAO/Clio was first characterized by \cite{Otten2017} and recently used by \cite{sutlieff2021high} for high-contrast imaging of a brown dwarf companion.

Seeing conditions over the course of the observation ranged from $0.39''$ to $0.53''$ with a median seeing of $0.44''$ measured by a differential image motion monitor (DIMM). Direct precipitable water vapor (PWV) measurements at the time of these observations are not available, but measurements of ground-level relative humidity indicate an estimate of 2.15~mm~PWV at the beginning of these observations, dropping to 1.92~mm by the end of them. The Sirius frames used for this study were taken from UT 05:58:17.11 to UT 08:55:32.35, encompassing 105$^\circ$ of sky rotation. Observations of \omigru{} for PSF calibration were taken shortly beforehand, with seeing $0.58''$ and approximate PWV of 2.0 mm. The conditions for both targets are summarized in Table~\ref{tab:conditions}.
% Seeing measurements were only available from 2015-11-29T05:38:38.03 to 2015-11-29T08:11:02.91 UT, but no notable excursions from this range were logged.

We observed Sirius using the vAPP-180 deg coronagraph with the [3.95] narrowband filter ($\lambda = 3.95\ \um{}$, $\Delta \lambda = 0.091\ \um{}$) to minimize radial smearing of the vAPP PSFs. We integrated for 500 ms each frame, obtaining a total of 12800 frames. We identified a stray light artifact impacting 1911 frames, leaving 10889 science frames to analyze encompassing $\approx90$ minutes of integration time.

The MagAO system was correcting 300 modes at 1~kHz, which resulted in residual wavefront error of approximately 100-150~nm RMS. At 3.95~\um{}, this corresponds to a Strehl ratio of $S \approx 96\%$, before including non-common-path aberrations. The field rotator was adjusted such that no stray light impacted the dark hole, and left in that position to enable the angular differential imaging (ADI) technique \citep{Marois2006}.

The Clio camera uses a legacy prototype Hawaii-I HgCdTe detector, sensitive out to 5 \um. Due to the extreme brightness of Sirius, the typical ``nodding'' procedure whereby the star is placed on alternating halves of the detector for sky estimation was not used. Instead, the star was periodically offset to a location completely off-chip, resulting in 1,220 total sky frames at intervals of 24 minutes.

Camera and MagAO configurations are summarized in Table~\ref{tab:observations}.

\section{Data reduction}

\begin{figure}
    \includegraphics[width=0.48\textwidth]{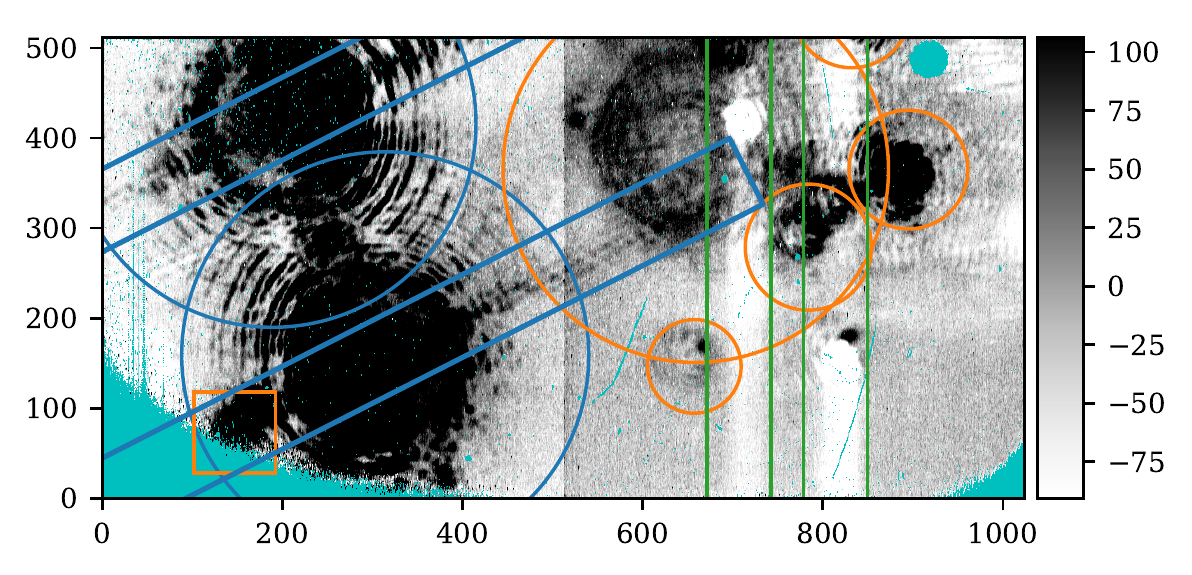}
    \caption{Single example frame after background subtraction, scaled to show structure. Bad pixels are cyan, the science PSFs and the cold stop diffraction spikes are outlined in blue, negative artifacts from amplifier crosstalk at a predictable separation from the star in green, and other stray light artifacts that can move over the course of the observation in orange. (Adapted from \citealt{Long2018}.)\label{fig:stray-light}}
\end{figure}

The extreme brightness of Sirius presented several hurdles to data reduction. Formerly unnoticed and negligible spurious reflections became non-negligible, as shown in Figure~\ref{fig:stray-light}. Not only did the PSF core saturate, but the first two diffraction rings of the PSF did as well. Short integration times translated into huge numbers of frames to incorporate in the reduction. In light of these challenges, we go into some detail about how we mitigated them.

\subsection{Sky background and stray light}\label{sec:bgsub}

The thermal infrared is challenging to observe from the ground due to the rapidly varying background illumination from the sky and non-cryogenic optics. %A typical ground-based IR observing strategy involves ``nodding'' the star from one half of the detector to the other, and using alternating frames as background references for the detector regions that were under the star in the preceding frame. Unfortunately, very few pixels on the Clio detector received negligible light from Sirius, regardless of which half it was placed on, rendering this technique unfeasible. Instead, the telescope was offset to place Sirius off-chip entirely to collect background frames at regular intervals.
Unlike stellar speckle noise, background counts are best modeled in the original pixel frame, before image alignment.
% We first attempted to subtract a scaled copy of the median of all the sky background frames, but the time series of RMS errors in reconstruction indicated this was not a good approximation. Instead,
We used the sky frames we recorded to construct a principal components basis. To model and subtract the background signal in a given science frame, we first identified pixels within a certain radius of the twin gvAPP-180 PSFs or a number of Clio-specific artifacts and excluded them (Figure~\ref{fig:stray-light}). \jladd{To mitigate impacts on sensitivity, only the small subset of pixels with negligible light from Sirius A were used to estimate the background in the reconstruction process.} We then performed a least-squares fit of the basis vectors to the pixels that remained, and used those coefficients to construct a background frame from the full un-masked principal component basis vectors \citep[][and independently in \citealt{Hunziker2018}]{Long2018}.

We held back 25\% of the sky frames in order to cross-validate the procedure, finding that a background model with \jladd{the first} 6 \jladd{principal} components reproduced background counts with RMS error 13 counts ($\approx 0.1\%$ of the average background level) when the fit excluded a mask representative of those for science frames. Although the final starlight subtraction is equally capable of subtracting the sky background, we found separating these steps into two stages improved our ability to create aligned cutouts and simplified later stages of the pipeline. \jladd{The error in background reconstruction of 13 counts amounts to $7\times 10^{-8}$ in contrast units for a star of Sirius' brightness, which we judge to be negligible.}

\subsection{PSF saturation and frame-to-frame variation}
\label{sec:scaling}

In half-second exposures with a coronagraph splitting the incoming beam in two, Sirius still saturated the PSF core completely, along with some of the diffraction rings.

\begin{figure}
    \includegraphics{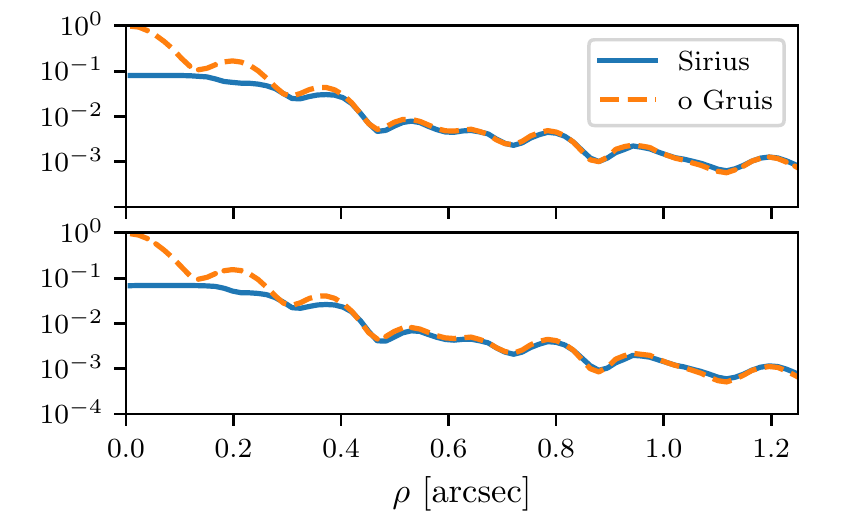}
    \caption{Radial profiles of the ``top'' (upper) and ``bottom'' (lower) vAPP PSFs (see Figure~\ref{fig:psf-saturation-images}) taken from median-combined data on Sirius and \omigru{}, normalized to 1.0 at $\rho = 0$ arcsec. The \omigru{} data allowed us to estimate the brightness of Sirius by fitting the wings of the PSF, and were used as an unsaturated PSF template when injecting fake planets.
    \label{fig:psf-saturation-profiles}}
\end{figure}

\begin{figure}
    \includegraphics{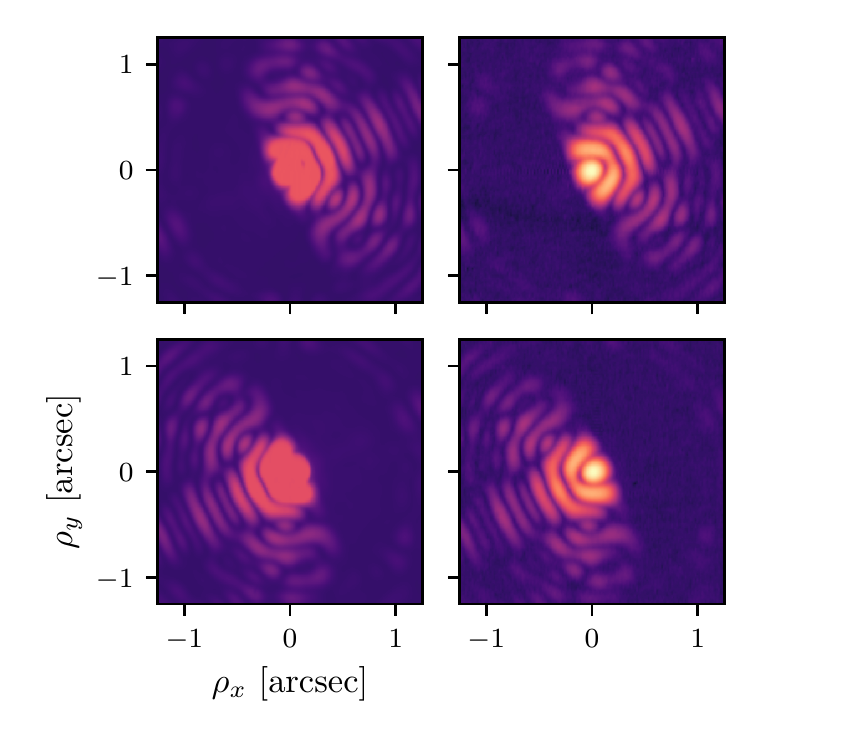}
    \caption{The first row shows the ``top'' vAPP median PSF, with saturated Sirius data on the left and \omigru{} data on the right. The second row shows the ``bottom'' PSFs. They have been scaled to relative units such that the peak of the unsaturated data are 1.0, and the saturated data are scaled to match the radial profiles in the unsaturated regions (see Figure~\ref{fig:psf-saturation-profiles}).
    \label{fig:psf-saturation-images}}
\end{figure}

In order to accurately model a companion, we needed to infer the missing flux by fitting a model to the unsaturated portions of the Sirius PSF in each of the two gvAPP-180 halves.
Shorter-integration Sirius data taken for calibration were also saturated, so we used a radial profile from stacked \omigru{} observations with the same filter and coronagraph configuration to serve as our unsaturated reference.
The \omigru{} observations were taken the same night (2015-11-29 UT) under similar conditions in terms of PWV and with the MagAO system correcting 300 modes.
The reported RMS residual wavefront error was approximately 100 nm RMS, for a Strehl ratio $S \approx 97\%$ before non-common-path errors.
In an ideal system, there would be no flux difference between the two gvAPP-180 halves, but \cite{Otten2017} noted that \jladd{one half of the split PSF was brighter than the other by several percent, and we also observed this in our data. As the PSFs are split by their left- or right-circular polarization, this discrepancy indicates a small amount of circular polarization is introduced by the instrument itself. As we observed the effect on multiple targets, we concluded the origin is instrumental, rather than any astrophysical polarization signal.}

We divided each frame into two cutouts centered on the complementary PSFs, which we aligned to sub-pixel precision. We established alignment to a known center of rotation by simulating a gvAPP-180 PSF without wavefront errors, using said PSF to align an unsaturated calibration target PSF (in this case, the \omigru{} template), and using this second, more realistic, PSF to align the Sirius frame cutouts. The result was two data cubes with dark-hole regions on opposite sides of the core, aligned to sub-pixel precision. We computed a radial profile of the template PSF, normalized to unit flux, and each frame's science PSFs.
At each frame, we fit the scale factor that matched the radial profiles in the unsaturated region best, giving us the total Sirius flux. A comparison of the profiles in the top and bottom PSFs is in Figure~\ref{fig:psf-saturation-profiles}. The effect of the ``lost'' flux is illustrated by Figure~\ref{fig:psf-saturation-images}.

Once in possession of two aligned data cubes, the amplitude difference between complementary gvAPP-180 PSFs was divided out using the per-frame scale factors.

\subsection{Starlight subtraction}

The brightness of Sirius required short integration times, which in turn led to a large number of frames containing distinct realizations of the system PSF. The KLIP algorithm of \cite{Soummer2012} leverages the difference between the number of realizations $n_{obs}$ and the number of random variables (i.e. pixels) $p$, under the assumption that $n_{obs} \ll p$, to reduce the size of the matrix to be eigendecomposed to obtain the Karhunen-L\`oeve basis. With $n_{obs} \sim 10,000$ and $p \sim 7,000$ for our analysis, this assumption no longer held. In fact, when $n_{obs} \simeq p$, the cost of forming the covariance matrix outweighed any time savings compared to a direct computation of the singular value decomposition (SVD) of the data matrix \citep{accelerateKlip}.

A common technique, adopted by \cite{Vigan2015} and many others, is to temporally bin the resulting data to reduce the dimensions of the problem. However, \cite{speckleLives} found that we should expect speckle lifetimes of roughly 20-100~ms on bright stars, well under even our 500 ms Sirius exposures. As such, we expected that capturing the information from every frame within an interval rather than temporally binning will result in improved starlight subtraction performance. This is, in fact, what we found for locations near the host star, as discussed in Section~\ref{sec:hyperparams}.

Initial attempts to apply the SVD downdate algorithm described in \cite{accelerateKlip} to accelerate the reduction of Sirius vAPP data proved insufficient. KLIP itself is vulnerable to self-subtraction in ADI data, as a planet PSF rotating through the field has enough overlap with itself from frame-to-frame that the resulting starlight subtraction process is able to remove or severely attenuate it. This is usually mitigated by use of a PSF reference star or an angular exclusion criterion, removing frames temporally adjacent to the one under analysis. However, applying an angular exclusion criterion undid some of the speed gains of the SVD downdate algorithm, as the inner SVD had to grow by a row and a column for each frame to be removed. In such a high frame-rate dataset, when probing locations of interest near the IWA large numbers of frames had to be excluded.

For this reason, we developed the PCA-Temporal (PCAT) algorithm detailed in Section~\ref{sec:pcat}. By operating on pixel time-series rather than frames, it is straightforward to mask a planet's path from the reference sample. This in turn means a single decomposition is required for all pixels (and therefore frames) to starlight-subtract.

\subsection{Fake planet injection and signal-to-noise ratio measurement}

After normalizing the PSF cubes with the profile fitting procedure in Section~\ref{sec:scaling}, we conducted ``fake planet'' signal injection/recovery tests with the \omigru{} template PSF as a realistic proxy for the unsaturated Sirius PSF. The paired PSFs of the gvAPP-180 required injecting the appropriate companion PSF template into each half of the data, as companions could rotate out of one dark hole and into another over the course of an observation.

To perform an injection-recovery test, a final focal plane location $(\rho, \theta)$ was selected (where $\rho$ is the separation from the center of the stellar PSF and $\theta$ is the position angle--PA--in degrees E of N). For each frame and each half, the template PSF was scaled by an amplitude $A$ and translated to $(\rho, \theta + \phi_i)$ in the focal plane (where $\phi$ is the negative of the derotation angle that places North-up and East-left for frame $i$) before adding it to the original image. To mitigate the influence of the injected planet signals on each other and the potential biasing of the estimated contrast, we only injected one companion at a time to measure the contrast limit in a location.

The starlight in each frame was subtracted following the algorithm in Section~\ref{sec:pcat}, and the residuals were derotated to place North-up East-left and combined as the pixel-wise median of the stack of frames.

The signal and noise samples were computed from the final image by simple aperture photometry in $\lambda / D$ diameter apertures, excluding the apertures on either side of the measurement location (i.e. known injected planet). The signal-to-noise ratio was then computed following the corrected equation for small-sample statistics from \cite{mawetSNR}.

\subsection{The PCA-Temporal algorithm}
\label{sec:pcat}

\begin{figure}
    \includegraphics[width=0.475\textwidth]{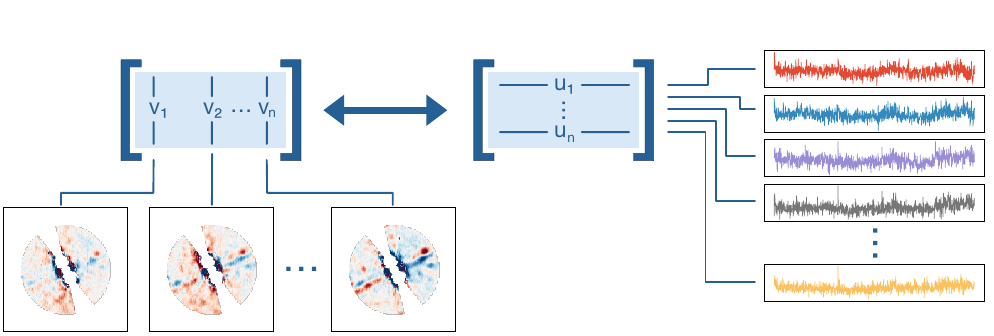}
    \caption{The two complementary decompositions of a hypothetical data matrix with one column for each frame. The matrix can be decomposed into left singular vectors, shown here as $v_1, v_2, \ldots, v_n$, or into right singular vectors $u_1, \ldots, u_n$. The interpretation of the left singular vectors, with the same number of entries as a single vector of pixels from an image, is as eigen-images of the pixel-to-pixel covariance, shown lower left. The analogous interpretation for the right singular vectors is as eigen-time-series from which each pixel's time series can be composed, shown at right. \label{fig:eigen-time-series}}
\end{figure}

Each focal plane pixel can be treated as a time-series that can be represented in a basis of eigen-time-series. \jladd{This technique is employed by Temporal Reference Analysis of Planets \citep[TRAP,][]{trap}, which also includes a constant term and a model light curve in the design matrix to be fit. However, the unique constraints of the paired vAPP PSF required development of a bespoke pipeline.}  By excluding pixels at the same separation as our injected companion from the basis determination, we avoid self-subtraction from overlapping companion flux. In essence, we have transposed the problem defined by \cite{Soummer2012} from eigenimages that describe the column space of a data matrix with one column per frame, to eigen-time-series that live in the row space as shown in Figure~\ref{fig:eigen-time-series}. In this way, we may use a single basis of eigen-time-series to starlight-subtract every frame, eliminating even the (reduced) per-frame cost of the SVD downdate. \jladd{Furthermore, by excluding an annulus of pixels from our reference set, we may reuse a single decomposition for multiple companion position angle values at a given separation, as the reference sets are identical. We found that a more selective mask covering only the notional companion's track, while better for including information on diametrically opposite speckles, also led to overfitting.}

To begin, a mask is optionally applied to select pixels from the $N$ frame sequence of images, and each frame's $p$ selected pixels are unwrapped into a vector $\mathbf{x}_{i}$, combined as the columns of a data matrix $\mathbf{X''}_{p \times N}$. Let $\tilde{\mathbf{x}}$ be a vector whose entries correspond to the median value of each row in matrix $\mathbf{X''}$.
The median-subtracted data are then $\mathbf{X'}_{p \times N} = \mathbf{X''}_{p \times N} - \tilde{\mathbf{x}}_{p \times 1}\mathbf{1}_{1 \times N}$.

To reduce the effect of brighter pixels on the decomposition, we divide each row $j$ of $\mathbf{X'}$ by a whitening factor given by the standard deviation $\sigma_j$ of that row. Let matrix $\boldsymbol{W}$ be a diagonal matrix $\mathbf{W} = \mathrm{diag}(\sigma_0^{-1},\ \dots,\ \sigma_p^{-1})$. The pre-processed observations matrix $\mathbf{X}$ is then given by

$$\mathbf{X}_{p \times N} = \mathbf{W}(\mathbf{X''}_{p \times N} - \tilde{\mathbf{x}}_{p \times 1}\mathbf{1}_{1 \times N})$$
\subsubsection{Special considerations for gvAPP-180 data}
When applying starlight subtraction algorithms to gvAPP-180 data, there are many possible ways to combine (or not combine) the data from the two halves into a final image. After experimentation, we found the best performance with fewest artifacts came from forming the image vector from the dark-hole pixels from one half, unwrapped into a vector and concatenated with the vector of dark-hole pixels from the other half such that the paired PSFs from a single frame map to a single observation vector $\mathbf{x}_i$. (This is equivalent to stitching the images beforehand, though this approach would also allow the use of the bright half of the PSF pixels in the reduction and using different masks for the estimation and the final combination steps.)

\subsubsection{Reference time-series selection}

Recall that to implement KLIP with ADI, one would exclude $a$ observation columns from $\mathbf{X}$ to make $\mathbf{R}_{p \times (N - a)}$, a matrix of reference observations assumed not to contain the signal of interest.
The analog in the time-domain is excluding time-series that would contain a planet ``transit'' signal.
The rate at which a planet PSF would pass over a detector pixel in ADI observations depends on its angular separation from the center of rotation (i.e. the host star). This means a planet-like signal ``transiting'' over any pixel at the same radius as our location of interest may cause it to be subtracted, if they are close enough in time. Therefore, we exclude $b$ of the pixel time-series in a ring of width $\Delta r = 2\ \lambda/D$, centered on the separation of the companion (whether it is one we injected or one we are trying to detect).
The reference data matrix is then of size $\mathbf{R}_{(p - b) \times N}$.

% The characteristic spatial noise scale of $\lambda / D$ requires a minimum of $(\mathrm{OWA} - \mathrm{IWA})$ independent runs of PCAT for a blind search, where OWA is the outer working angle in $\lambda / D$ and IWA is the inner working angle. For each step in $\lambda / D$, a different annular mask would exclude the pixels under it from the reference sample. Each of those steps would require decomposition of a matrix of reference observations, but the ability to reduce all  within a run with the same basis more than offsets this additional computation time.

\subsubsection{Eigen-time-series computation}

Having selected the reference pixel time-series vectors to form $\mathbf{R}$, the algorithm proceeds with a singular value decomposition (SVD) to obtain the left and right singular vectors as well as their associated singular values. We use SVD to mean the ``economy'' SVD that omits the decomposition of the null space of the matrix. The SVD allows us to decompose $\mathbf{R}$ into

$$\mathbf{R} = \mathbf{U}_{(p - b) \times K}\boldsymbol{\Sigma}_{K \times K}(\mathbf{V}_{N \times K})^T$$

where $K = \min((p - b), N)$. To prevent overfitting and improve starlight subtraction, we retain the $k < K$ column vectors in the decomposition corresponding to the greatest singular values. (If $k \ll K$, this may allow the use of faster algorithms that find the partial SVD.) The decomposition is now only approximate, and given by

$$\mathbf{R} \approx \mathbf{\tilde{R}} = \mathbf{\tilde{U}}_{(p - b) \times k}\boldsymbol{\tilde{\Sigma}}_{k \times k}(\mathbf{\tilde{V}}_{N \times k})^T.$$

The question of the optimal choice of $k$ is addressed in Section~\ref{sec:hyperparams}. The matrix $\mathbf{\tilde{V}}$ defined above gives us the eigen-time-series with which we will model each pixel's evolution.

\subsubsection{Subtraction of estimated starlight}

Up until now, we have referred to observation column vectors $\mathbf{x}_i \in \mathbb{R}^p$. Now we operate on the rows of $\mathbf{X}$, which we will call $\mathbf{y}_j \in \mathbb{R}^N$. For every pixel time series we use from the original data cube, we proceed to project it into the basis of eigen-time-series $\mathbf{\tilde{V}}$ and reconstruct it in terms of the first $k$ eigen-time-series:

$$\mathbf{\tilde{y}}_{j} \approx \mathbf{\tilde{V}} ((\mathbf{\tilde{V}}^T)_{k \times N}\, \mathbf{y}_{j})$$

The residual noise plus companion signal (if any) is a vector $\mathbf{s}_j \in \mathbb{R}^N$ given by $\mathbf{s}_j = \mathbf{y}_j - \mathbf{\tilde{y}}_j$.

By repeating this procedure for all $p$ rows of $\mathbf{X}$, we obtain the starlight-subtracted signal $\mathbf{S}$ whose rows are $\mathbf{s}_1$, $\mathbf{s}_2$, \dots, $\mathbf{s}_p$.

$$\mathbf{S}_{p \times N} = (\mathbf{X}^T - \mathbf{\tilde{V}} (
    \mathbf{\tilde{V}}^T \mathbf{X}^T
))^T$$

To reverse the whitening transformation, we use $\mathbf{W}^{-1}$ and obtain the final starlight subtracted data $\mathbf{D}$:

$$\mathbf{D} = \mathbf{W}^{-1}_{p \times p} \mathbf{S}_{p \times N}.$$

The entries of the columns of $\mathbf{D}$ may then be mapped back into pixel locations in the original data cube to construct a cube of residuals and further post-processed (e.g. by derotating and stacking images for ADI).

\subsection{Optimization of PCAT hyperparameters}
\label{sec:hyperparams}

Due to the unique structure of the gvAPP-180 data, the optimal number of modes to subtract $k$ varied with both separation and position angle. Additionally, when calibrating the SNR=5 contrast floor, we observed that injecting and recovering a signal of amplitude $A$ producing SNR $S \gg 5$ would predict a larger value $A_\mathrm{SNR=5} = (5 A) / S$ for the minimum contrast detectable at SNR=5 than an injection-recovery test closer to the true SNR=5 amplitude. \jladd{Or, to put it another way, calibrating an accurate minimum contrast value depends on injecting and recovering a signal close to the minimum amplitude that is detectable at a good SNR, making this an iterative process.} Calibrating the SNR=5 contrast floor therefore required tuning of two hyperparameters for every location probed: injected signal amplitude, and number of PCAT modes. The width of the annular mask excluding pixels from the reference sample was fixed at $b = 16$ pixels ($\approx 2 \lambda / D$).

The hyperparameter tuning problem is an area of ongoing research in the machine learning literature (see e.g. \citealt{yangHyperparameterOptimizationMachine2020} for a review), with multiple techniques to efficiently explore the parameter space at our disposal. Since each parameter evaluation is comparatively expensive, we chose Bayesian optimization \citep{snoekPracticalBayesianOptimization2012}, descended from the ``kriging'' technique (also known as Gaussian process regression) initially developed to predict the profitability of mining gold deposits from a small number of bore-hole samples \citep{krige1951statistical}.

\begin{figure}
    \includegraphics{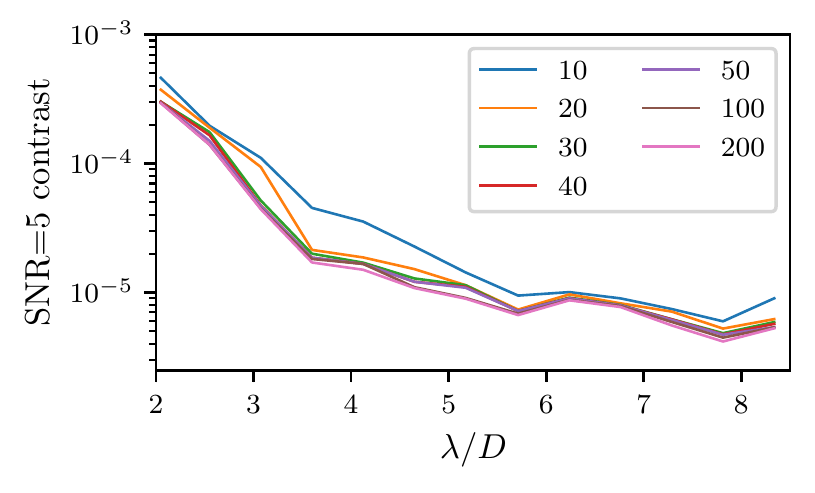}
    \caption{\jladd{Median SNR=5 contrast vs. separation shown at 10, 20, 30, 40, 50, 100, and 200 iterations of hyper\-parameter exploration by the Bayesian optimization algorithm. The Bayesian optimization procedure rapidly converges to near-optimal parameters.}\label{fig:show-convergence}}
\end{figure}

The optimization process attempts to learn a relationship between the number of PCAT modes, $k$, injected signal amplitude, $A$, and the resulting recovered SNR $S$ at some location. The objective function we chose to maximize is $$f(A, S) = -S \log_{10} A.$$ This depends only on the recoverability of the injected signal, and does not attempt to optimize the SNR from reducing the no-injection dataset. A separate forthcoming publication (Long \emph{et al.}, in prep.) will discuss the particulars of our approach to hyperparameter optimization and the distributed computing infrastructure developed for it. \jladd{The optimization process was run for 200 iterations of selecting parameters, injecting a signal, reducing the data with said parameters, and measuring the recovered signal. Figure~\ref{fig:show-convergence} shows the rapid convergence of the optimizer, with only marginal contrast gains after 50 iterations.}

\begin{figure}
    \includegraphics{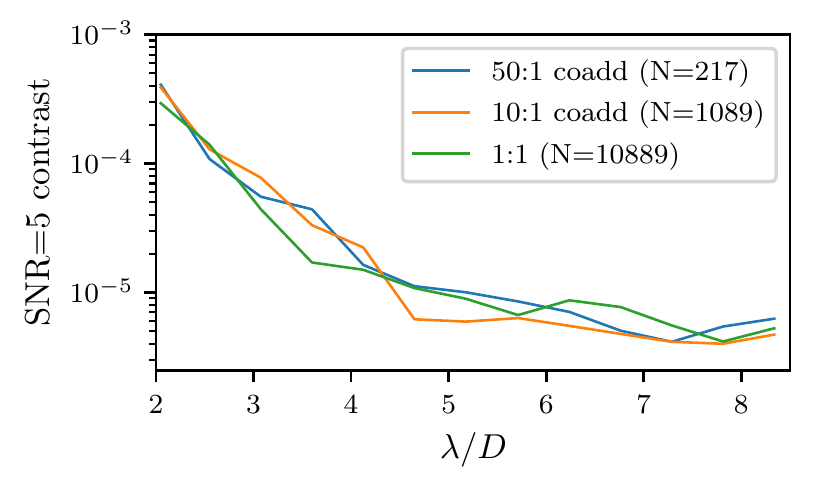}
    \caption{\jladd{Median SNR=5 contrast vs. separation for three different levels of co-adding.} Within 4~$\lambda / D$, the median contrast degrades with temporal downsampling by coadding. This relationship turns over, and by 5~$\lambda / D$ the 10:1 downsampled data cube actually provides better contrast.\label{fig:compare-coadds}}
\end{figure}

In recognition of the difference in noise statistics between points in the speckle-dominated region close to the star, and background-limited points further out, we did compute contrast limits with varying amounts of co-adding (shown in Figure~\ref{fig:compare-coadds}). Retaining every frame in full temporal sampling produced better contrast limits at small separations, while performance at larger separations benefitted somewhat from co-adding temporally adjacent frames.

\begin{figure}
    \includegraphics{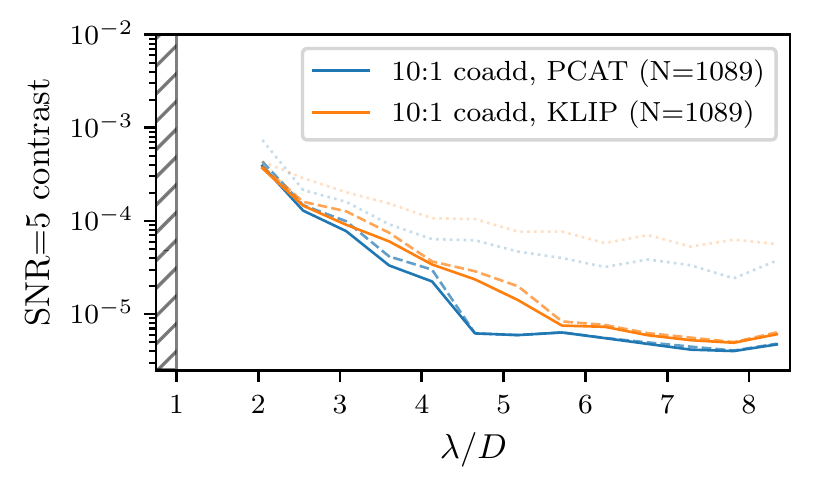}
    \caption{Behavior of optimized median SNR=5 contrast at each radius value with increasing number of optimization iterations, comparing PCAT to our implementation of KLIP. Both algorithms were optimized using the same Bayesian optimization procedure, with the median SNR=5 contrast shown after 10 iterations as a dotted line, after 100 as a dashed line, and after 200 as a solid line.\label{fig:compare-klip}}
\end{figure}

We validated the performance of PCAT by comparison with our implementation of KLIP, with the same optimization scheme. We optimized the 10:1 coadded reduction with KLIP following the same procedure as PCAT, and show the results in Figure~\ref{fig:compare-klip}.

\begin{figure}
    \includegraphics{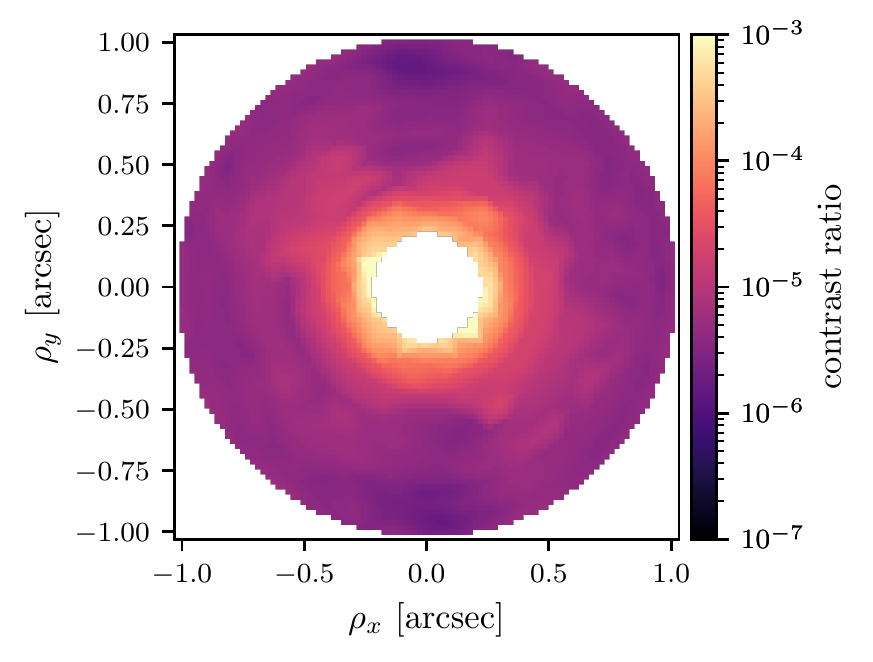}
    \caption{Contrast map of the faintest signal detectable at SNR=5, expressed as a ratio to the host star flux.\label{fig:contrast-map}}
\end{figure}

\jladd{
    To compute the final contrast surface, we collected all combinations of parameters evaluated for a particular focal plane location where the injected signal was recovered with at least a signal-to-noise ratio of $S \ge 8$. This cutoff was chosen arbitrarily to limit the analysis to points with relatively trustworthy estimates of the SNR=5 contrast level $A_\mathrm{SNR=5} = 5A/8$. The parameter combination with the smallest  $A_\mathrm{SNR=5}$ was saved, and those values define the surface in Figure~\ref{fig:contrast-map}. The $k$ modes, injected amplitude $A$, and temporal downsampling factor $n$ were allowed to vary, though we only evaluated fixed steps of $n \in \{1, 10, 50\}$ due to computing resource constraints. The spatial distributions of these parameters are shown in Appendix~\ref{sec:hyperparam-distributions}.
}

\section{Detection limits}

% \begin{figure}
% \includegraphics{contrast-angular.pdf}
% \caption{The line indicates the median contrast level for which we would detect a source at SNR=5, taken over the contrast levels for all position angles at that separation. The variation of contrast with position angle is captured by the shaded gray region on either side of the line.\label{fig:contrast-angular}}
% \end{figure}

% \begin{figure}
% \includegraphics{contrast-lamd.pdf}
% \caption{Exactly the data shown in Figure~\ref{fig:contrast-angular} but scaled to the wavelength and diameter of the telescope used.\label{fig:contrast-lamd}}
% \end{figure}

\begin{figure}
\includegraphics{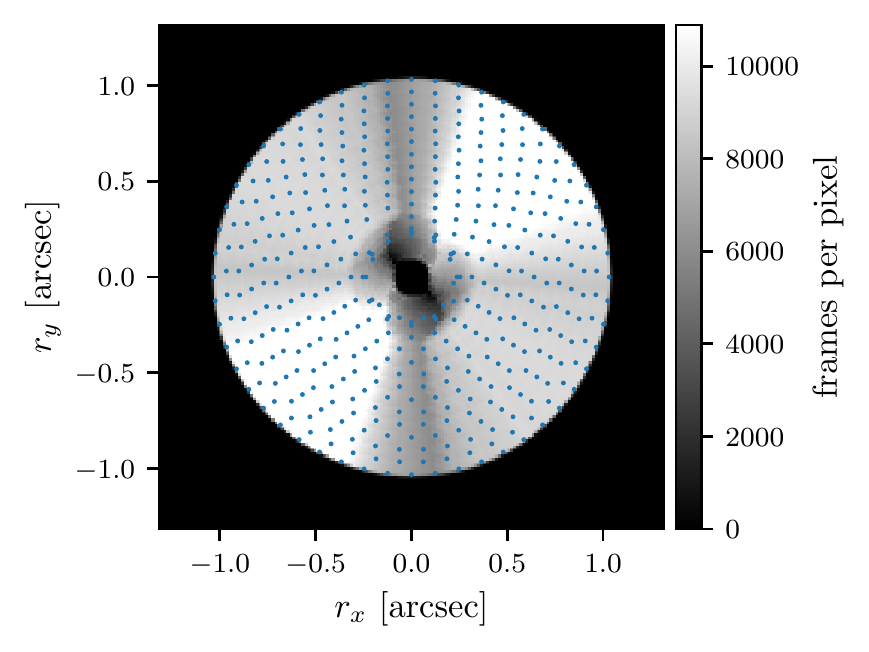}
\caption{The pattern of saturated pixels in each complementary half of the gvAPP-180 PSF, combined with the field rotation, leads to a varying number of frames from the original science data contributing to the value of each final pixel. This picture shows the result of rotating the data masks (set to 1.0 where data are kept) and summing along the time axis to visualize this effect. The overlaid points represent locations at which we injected and recovered a planet signal to calibrate our contrast limits.\label{fig:coverage}}
\end{figure}

\begin{figure}
    \begin{center}\includegraphics{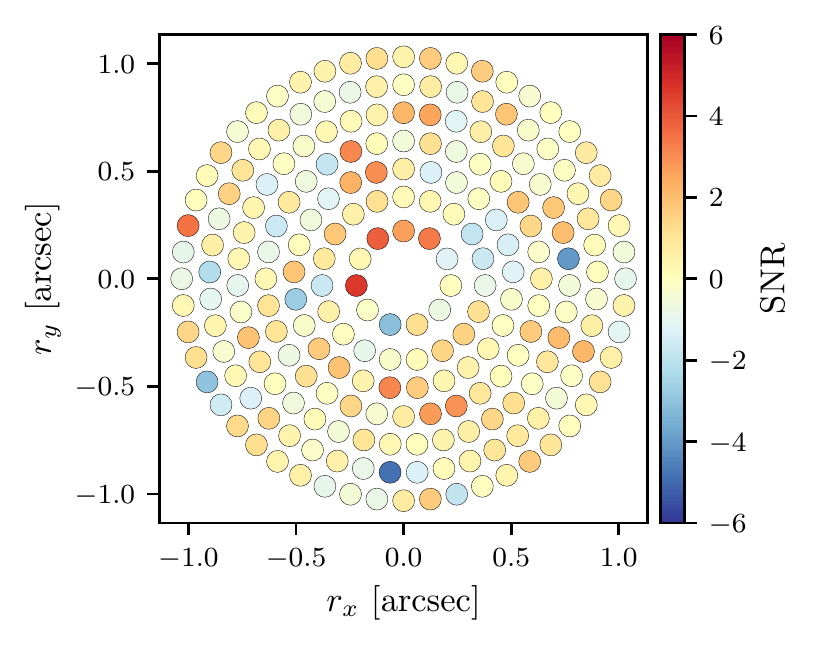}\end{center}
    \caption{SNR map of the final de-rotated focal plane. To calculate the detection map, the optimized mode fraction $k$ for a particular location $(\rho, \theta)$ is taken from the optimization process and applied to reducing the original data \emph{without} an injected companion. The resulting SNR, measured in $\lambda / D$-spaced apertures and computed with the correction from \citealt{mawetSNR}, is shown in this figure. We found no SNR $\ge 4$ signals to report, and the lower-SNR signals we inspected did not behave like a companion (moving with changing $k$, or disappearing) which points to a local maximum of the optimization process producing a spurious detection.\label{fig:detection}}
\end{figure}

\begin{figure*}
    \includegraphics{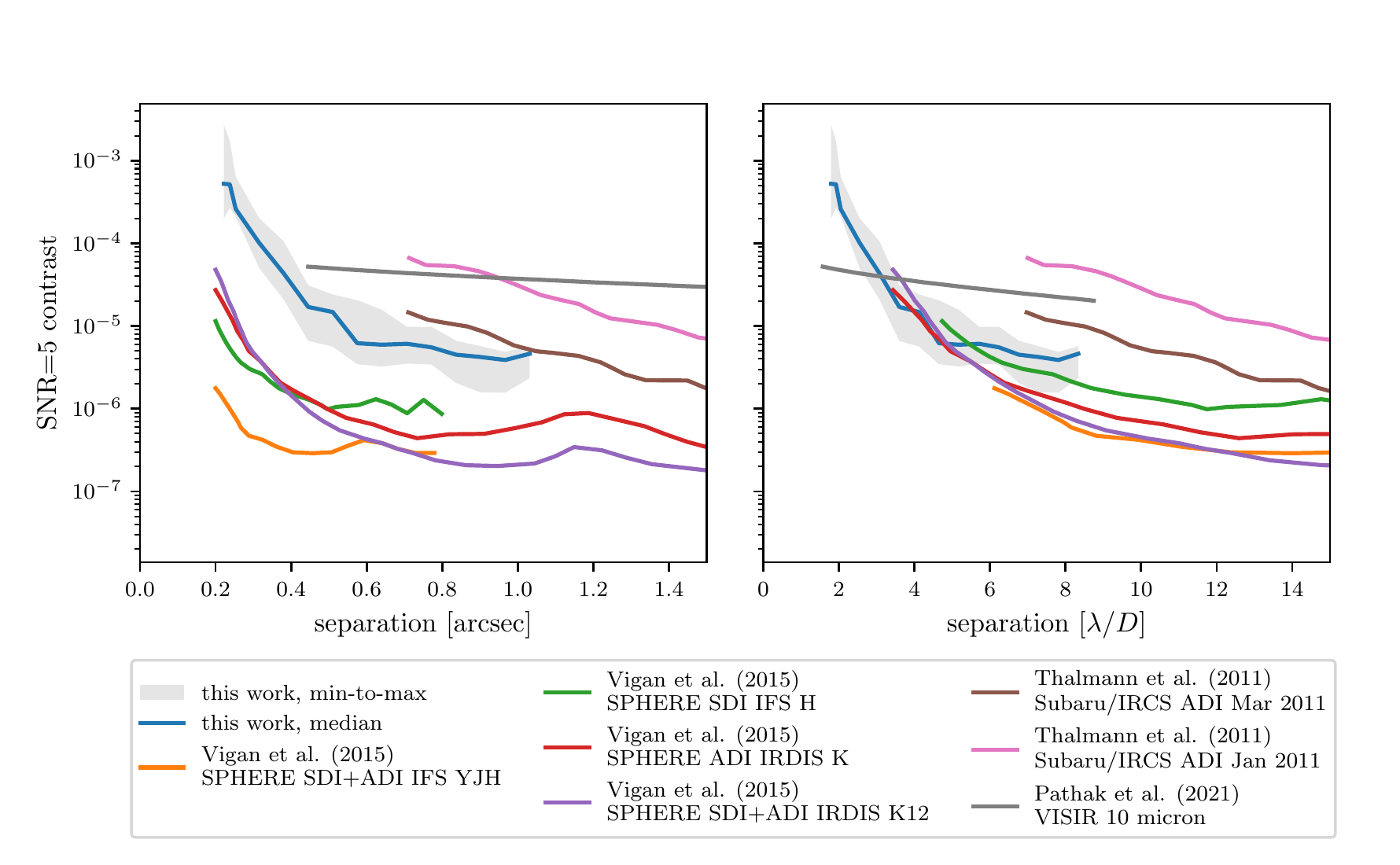}
    \caption{The line labeled ``this work, median'' indicates the median contrast level for which we would detect a source at SNR=5, taken over the contrast levels for all position angles at that separation. The variation of contrast with position angle is captured by the shaded gray region on either side of the line. \emph{(left)} The contrast levels as a function of separation in arcseconds with \cite{Thalmann2011,Vigan2015} and \cite{pathakHighcontrastImagingTen2021} reported curves for comparison. \emph{(right)} The contrast levels scaled into $\lambda / D$ units to more effectively compare inner working angle across wavelengths and telescopes. When scaled by the telescope diameter and wavelength, the ability of the gvAPP-180 to work at extremely small inner working angles is apparent. \label{fig:contrast}}
\end{figure*}

Using the procedure described above, we calibrated the minimum brightness of an SNR=5 signal at a set of focal plane locations arranged in rings around the center of the host star PSF (Figure~\ref{fig:coverage}). \jladd{To obtain the detection map in Figure~\ref{fig:detection}, we took those parameter combinations that produced the best $A_\mathrm{SNR=5}$, and applied them to reduce the pristine data (with no injected signals).} Our gvAPP-180 coronagraph allows for remarkably deep contrast at close separations, as shown in Figure~\ref{fig:contrast}. When scaled to the wavelength and diameter of the telescope used, the smaller IWA of the gvAPP-180 is apparent (Figure~\ref{fig:contrast}, right).

It is important to note that the radially averaged contrast limit curve does a poor job capturing the variation in coverage and achieved contrast floor with position angle, shown in Figure~\ref{fig:contrast-map}. Due to the geometry of the coronagraph, masked region, and sky roation, the final coverage in terms of the number of science frames contributing to each pixel varies as a function of position angle---especially at the innermost separations probed (Figure~\ref{fig:coverage}).
%(Later gvAPP designs reduced the ``dead zone'' between the complementary dark hole regions, and the gvAPP-360 design creates a symmetric dark hole region with a single PSF.)

\subsection{Interpretation as a companion mass limit}

Predicting the properties of planets that one would be sensitive to depends on choosing a model for planet evolution and spectra. We adopt the latest ``Bobcat'' models in the ``Sonora'' series of \cite{marleyBobcat}, which span a range of $0.5 < M / M_J < 84$ in planet mass and 1~Myr~$< t <$~15~Gyr in age. They model planets and substellar objects down to $T_\mathrm{eff} = 200$ K with associated evolutionary tracks, high-resolution simulated spectra, and different metallicities. We perform our own synthetic photometry for this analysis using the published spectra.

The Bobcat models are published for $[M/H] \in \{-0.5, 0.0, +0.5\}$. \cite{Bond2017} models the evolution of Sirius A, reporting a slight metal deficiency relative to solar abundances and that $[Fe/H]$ of -0.13 or -0.07 reproduces observations. We adopt the $[M/H] = 0.0$ Bobcat models as a conservative option, because lower metallicity models have higher [3.95] flux. As an example, the absolute [3.95] magnitude of 14 corresponds to masses of 7.5, 8.5, and 8.8 $M_J$ in the $[M/H]$ -0.5, 0.0, and +0.5 model suites, respectively.

\label{sec:irradiation}
\subsubsection{Incorporating the effects of irradiation from Sirius~A}

\begin{figure}
    \includegraphics{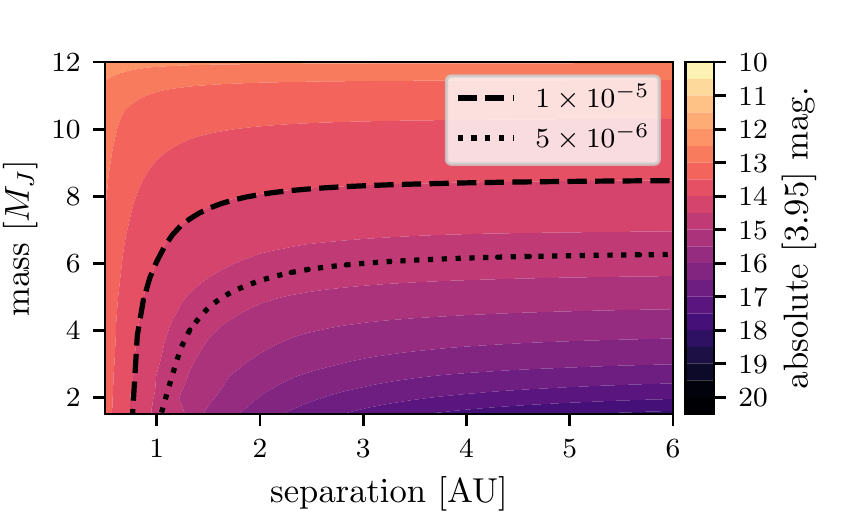}
    \caption{The masses detectable at a host/star contrast of $10^{-5}$ and $5 \times 10^{-6}$ are shown overlaid on a contour plot of planet absolute [3.95] magnitudes predicted by the Bobcat models for $t=242$ Myr and $T_\mathrm{eff} = 9910$ K for Sirius. \label{fig:planet-models-mag-to-mass}}
\end{figure}

\jladd{Instellation} from Sirius A would be a major contributor to the effective temperature of a giant planet at a small separation from the star. \cite{malesDirectImagingExoplanets2014} showed that the mass/separation/absolute magnitude relationship depends only weakly on mass at small separations, effectively making lower mass planets brighter and thus more accessible to current-generation instruments. To account for this in our mass limits, we computed an equilibrium temperature
\begin{equation}
    T_{eq} = T_\mathrm{Sirius} \sqrt{\frac{R_\mathrm{Sirius}}{2 a}} (1 - A_B)^{1/4}
\end{equation}
where $A_B$ is the Bond albedo and $a$ is the separation between star and planet. \cite{marleyReflectedSpectraAlbedos1999} found that cloud-free planets with incident flux from an A star have a mass- and temperature-dependent albedo ranging from $0.39 \le A_B \le 0.43$ in their simulations. That range extends to $0.39 < A_B < 0.93$ when considering different cloud models. For this analysis, we adopt $A_B = 0.5$ for consistency with Jupiter \citep{liLessAbsorbedSolar2018}.

The giant planets we might expect to find will glow with the heat of their formation and cool slowly over time to their equilibrium temperatures. We approximate the contribution to the equilibrium temperature $T_\mathrm{eq}$ from incident light by summing the planet luminosity \begin{equation}
    L_\mathrm{eff} = L_\mathrm{evol} + L_\mathrm{eq}
\end{equation}
where
\begin{eqnarray}
    L_\mathrm{evol} &= \sigma_\mathrm{sb} 4 \pi R_\mathrm{planet}^2 T_\mathrm{evol}^4\\
    L_\mathrm{eq} &= \sigma_\mathrm{sb} 4 \pi R_\mathrm{planet}^2 T_\mathrm{eq}^4
\end{eqnarray}
and $T_\mathrm{evol}$ is the evolutionary temperature from the Bobcat isochrones. Substituting the Stefan-Boltzmann law for $L_\mathrm{eff}$ gives us the following relationship for $T_\mathrm{eff}$ of the irradiated planet

\begin{eqnarray}
    \label{eq:teff}T_\mathrm{eff}^4 &= T_\mathrm{evol}^4 + T_\mathrm{eq}^4.
    % T_\mathrm{eff} &= \left[T_\mathrm{evol}^4 + T_\mathrm{eq}^4\right]^{1/4}
\end{eqnarray}

This modified relationship in the specific case of Sirius and the [3.95] filter is shown in Figure~\ref{fig:planet-models-mag-to-mass}. We do not claim to have captured any of the subtleties of the evolution of an irradiated planet, only to have chosen an appropriate $T_\mathrm{eff}$ subject to what we know about the minimum equilibrium temperature for an assumed separation and albedo value. Because of the way the Bobcat models were initialized, models with very high $T_\mathrm{eff}$ for very low masses were not available. Therefore, the effective minimum modeled mass for our analysis is $1\,\mathrm{M}_J$.

\subsubsection{Magnitude-to-mass relationship}
To obtain the relationship between the modeled masses and their fluxes in the $[3.95]$ filter, we assumed a system age of 242~Myr based on the determination of Sirius A's age in \cite{Bond2017}. For every modeled companion mass, $T_\mathrm{eff}$ and surface gravity were computed from the evolution tracks subject to the modification described above. We then interpolated a spectrum from the Bobcat grid, and computed a [3.95] absolute Vega magnitude for such an object.

To compute the synthetic photometry, we adopted the latest Vega model from HST CALSPEC\footnote{{\tt alpha\_lyr\_mod\_004.fits}} \citep{calspec} and an atmosphere with 2 mm of PWV and $\sec(z) = 1.15$. The modeled atmosphere retrieved from ESO SkyCalc \citep{jonesAdvancedScatteredMoonlight2013,nollAtmosphericRadiationModel2012} represents La Silla observatory, quite close to Las Campanas where these data were taken, and its slightly lower altitude means its extinction is slightly pessimistic compared to actual.

\begin{figure*}
\includegraphics{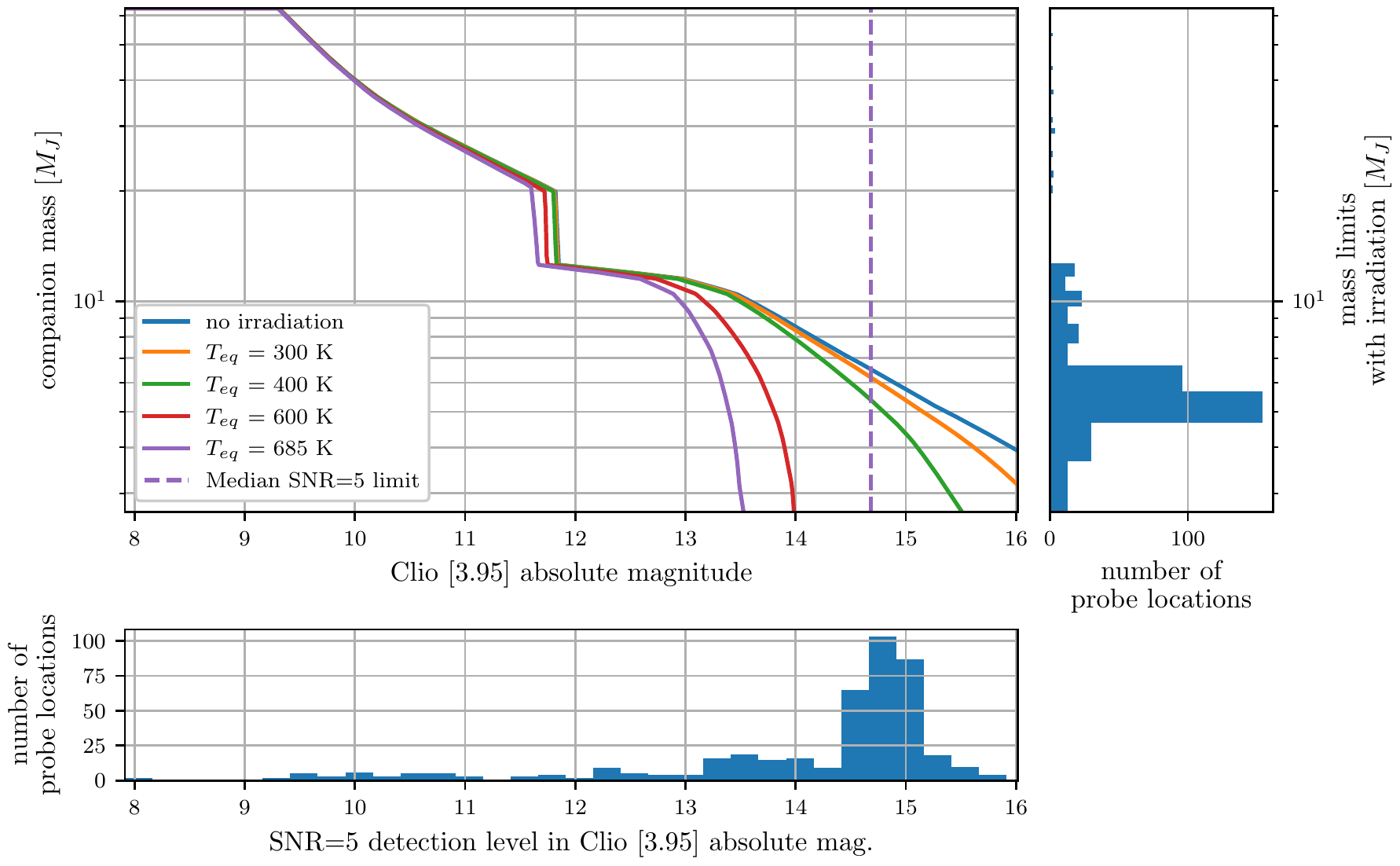}
\caption{Correspondence of [3.95] absolute magnitudes to masses assuming $A_B = 0.5$. {\it (main panel)} The relationship between magnitude and mass is plotted at age 242 Myrs and several different equilibrium temperatures to illustrate its dependence on $T_\mathrm{eq}$. The $T_\mathrm{eq} = 685$ K line corresponds to the relationship at the smallest separation probed at the IWA. The dashed line indicates the median absolute magnitude value of the contrast floor across all focal-plane locations. The apparent discontinuity around 12 $M_J$ is where the relationship becomes double-valued as deuterium burning begins, and we take the lower value of the two. {\it (lower panel)} Histogram of SNR=5 detection levels reached at the probe locations shown in Figure~\ref{fig:coverage} in 0.25 mag bins.{\it (right panel)} Histogram of probe locations reaching particular mass sensitivity limits in 1 $M_J$ bins. \label{fig:mag-to-mass}}
\end{figure*}

With a [3.95] magnitude for each mass computed,
we inverted the relationship. This relationship is shown for a selection of $T_\mathrm{eq}$ values in
Figure~\ref{fig:mag-to-mass}. Where the relationship is double-valued for masses around the minimum mass for deuterium burning, we made it monotonic by taking the minimum mass value for that magnitude, resulting in the apparent discontinuity around $m_{[3.95]} = 11.5$.

\subsubsection{Mass limits}

\begin{figure*}
\includegraphics{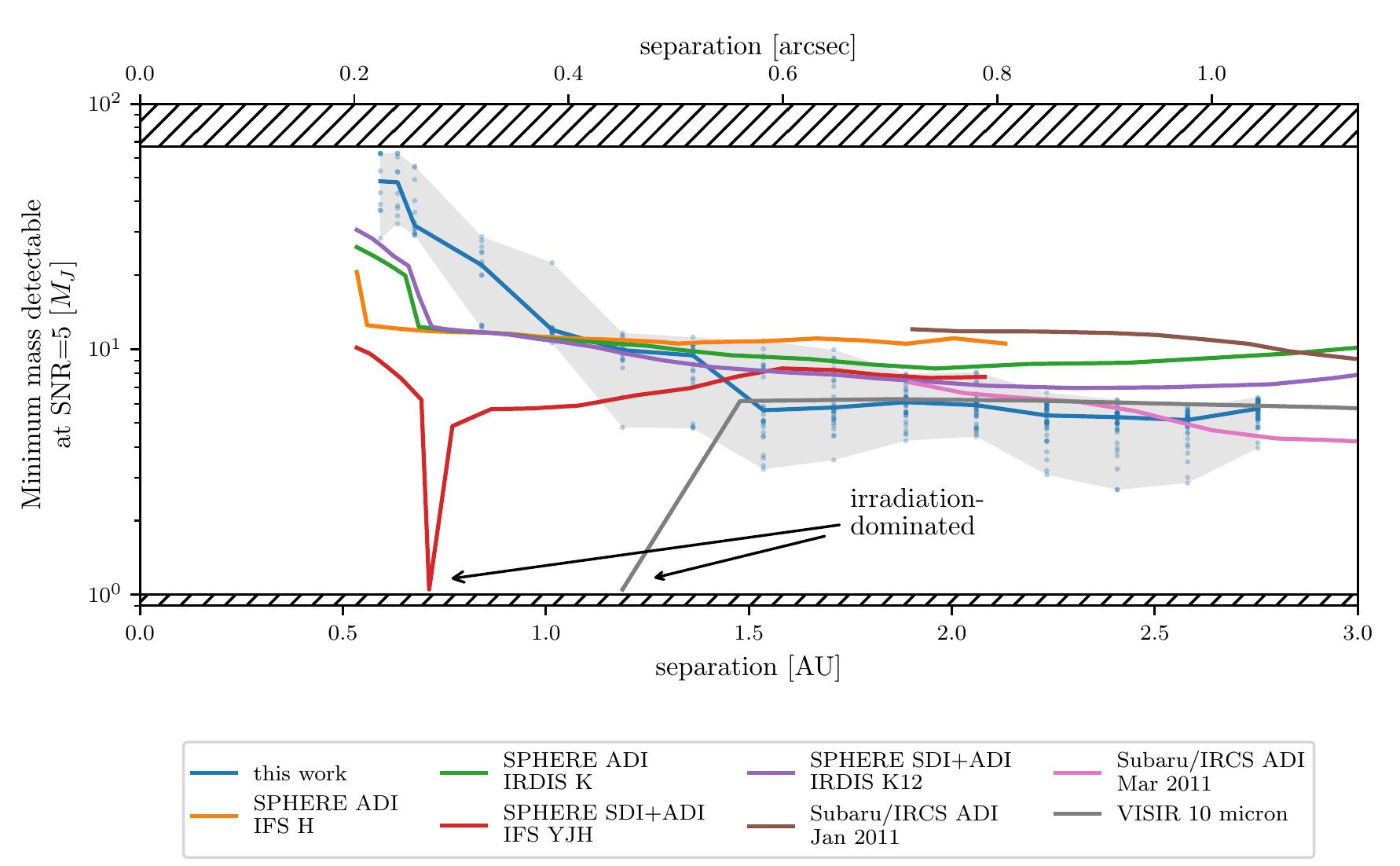}
\caption{The minimum mass detectable at SNR=5 as predicted from our contrast levels using the \cite{marleyBobcat} models. Blue line and points are from this work. The contrast curves reported by \citealt{Thalmann2011} (Subaru/IRCS), \citealt{Vigan2015} (SPHERE), and \citealt{pathakHighcontrastImagingTen2021} (VISIR) have been re-analyzed using the Bobcat models for spectra and evolution, incorporating a $T_\mathrm{eq}$ correction as described in Sec.~\ref{sec:irradiation} with $A_B = 0.5$. The shaded region encompasses the minimum to maximum mass across all sampled points at that radius. Points at 1~$M_J$ are likely overestimates, as that is the lower limit of the range of the model grid. (Shown as a cross-hatched region for out-of-bounds masses.)\label{fig:mass-limit}}
\end{figure*}

\begin{figure}
\includegraphics{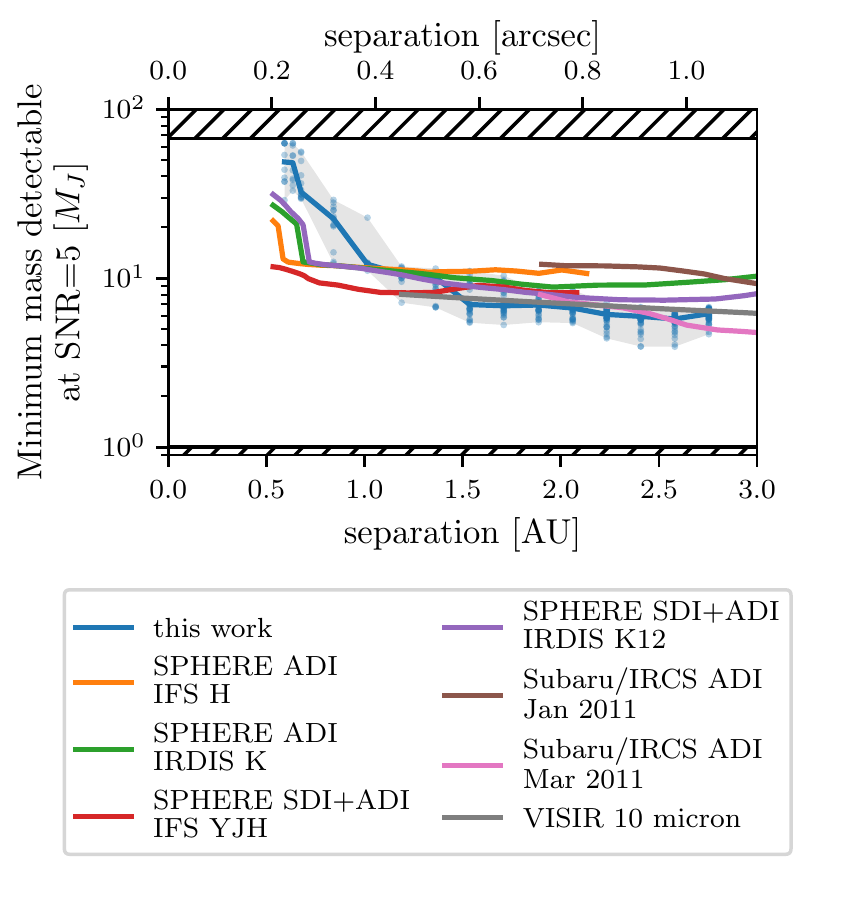}
\caption{The minimum mass detectable at SNR=5 as predicted from our contrast levels using the \cite{marleyBobcat} models, without irradiation ($A_B = 1.0$). Blue line and points are from this work.\label{fig:mass-limit-no-irr}}
\end{figure}

The minimum mass we would detect at SNR=5 is shown as a function of separation in Figure~\ref{fig:mass-limit}. The overlaid scatter points and shaded region illustrate the variation in that mass limit as a function of angle at that separation, which is a consequence of the rotation of the dark hole region, changes in conditions over the course of the observation, etc.

\begin{figure}
    \includegraphics{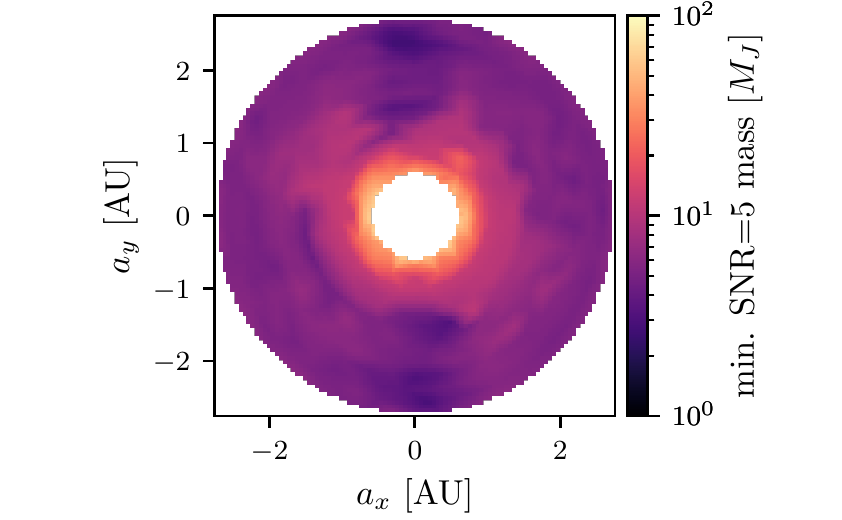}
    \caption{Visualization of the mass limit. This minimum mass is detectable when the separation is entirely in the plane of the sky, in other words, when the planet is experiencing the maximum irradiation for that projected separation. Since the irradiation depends on the actual separation, while the contrast limit depends on the projected separation, it is easier to interpret these results through completeness plots.\label{fig:mass-map}}
\end{figure}

To visualize the two-dimensional variation in the mass limit, when considering a point at angular coordinates $(\rho, \theta)$ in the focal plane, $a = d_\mathrm{Sirius} \tan \rho$ is used as the separation for the purpose of computing a modified magnitude-to-mass relationship. The minimum detectable masses at each position are shown in Figure~\ref{fig:mass-map}, adopting a Bond albedo $A_B = 0.5$.

While this is sufficient to exclude a planet at $(\rho, \theta)$ of that mass (or greater) and that albedo at separation $a$, it cannot necessarily exclude a planet of that mass but a greater separation that happens to fall on $(\rho, \theta)$ in the focal plane when viewed from Earth. A planet at that projected separation, but far enough away to experience negligible irradiation, is equivalent to a planet with an albedo $A_B = 1.0$. The mass limits under this assumption are given in Figure~\ref{fig:mass-limit-no-irr}.

Because this relationship depends on the projected separation (for the contrast limit) and the true separation (for the $T_\mathrm{eq}$), it is more convenient and informative to fold these variables into the completeness analysis.

\subsection{Completeness limits}

\begin{figure}
    \includegraphics{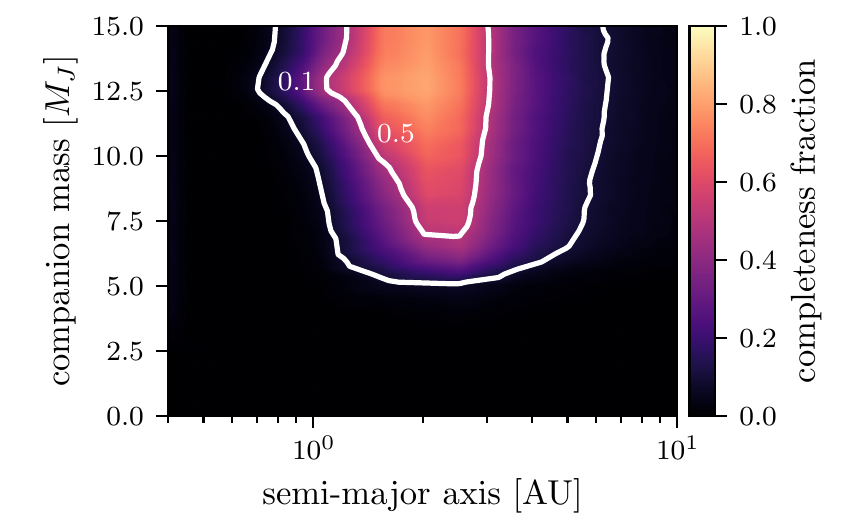}
    \caption{The fraction of planets on randomly drawn orbits around Sirius A that would be detectable in these observations, as a function of semi-major axis and mass. (Note that the separation axis is scaled differently from that in Figure~\ref{fig:mass-limit}.)\label{fig:completeness-clio}}
\end{figure}

\begin{figure}
    \includegraphics{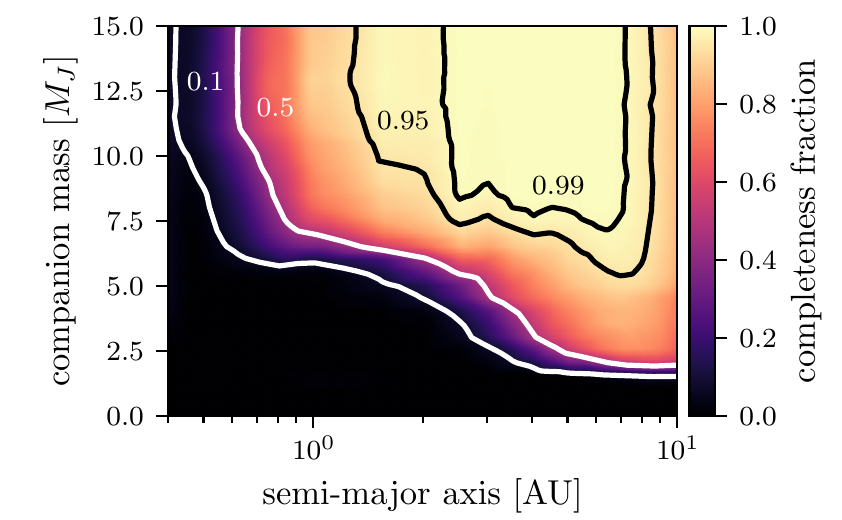}
    \caption{The fraction of planets on randomly drawn orbits around Sirius A that would be detectable in these observations and/or a past study of Sirius A, as a function of semi-major axis and mass. When combining the results of all four studies, at least one study would have detected a mass 9 $M_J$ planet in the 2.5--7 AU range for 99\% of trials.\label{fig:completeness}}
\end{figure}

Our sensitivity is a function of the contrast floor (itself a function of position in the focal plane), companion mass, semi-major axis, and distance from the host star at the moment of observation. To capture this in our completeness analysis, we drew $10^5$ random orbital elements for a grid of masses and semi-major axis values. The eccentricities are drawn from a linearly descending prior given in \cite{nielsenGeminiPlanetImager2019}.

Projecting the position of the notional companion into the plane of the sky gives us the separation and position angle we would observe, while the orbit determines the actual separation and hence the irradiation-affected effective temperature. The assumed age and metallicity, plus the companion mass and the projected and actual separation, then allow us to compute a [3.95] magnitude for the companion. In Figure~\ref{fig:completeness-clio}, the fraction of the trial orbits that result in [3.95] magnitudes brighter than our SNR=5 detection level give the completeness at each mass and semi-major axis point.

Using the SNR=5 contrast vs. separation curves of each of the studies, as well as their dates of observation, we were able to assess whether a randomly drawn orbit would have been detected in at least one of the studies. We converted the other studies' contrast limits into mass limits following the same procedure as our own, assuming radial symmetry. For each point in (mass, semi-major axis) space, we drew $10^5$  sets of orbital elements. Each was converted into a set of positions along the orbit using the observation dates of the studies.

We used the resulting true separations and projected separations to generate synthetic photometry in the bandpasses corresponding to each published contrast curve. As before, irradiation from the primary was taken into account. If any of the studies would have detected the planet at the corresponding position for their observation date, we counted it as detectable in the final combination of the completeness grid. The overall completeness from all four studies as a function of mass and semi-major axis is shown in Figure~\ref{fig:completeness}.

\section{Considering a second epoch of formation triggered by Sirius B}

\begin{figure}
    \includegraphics{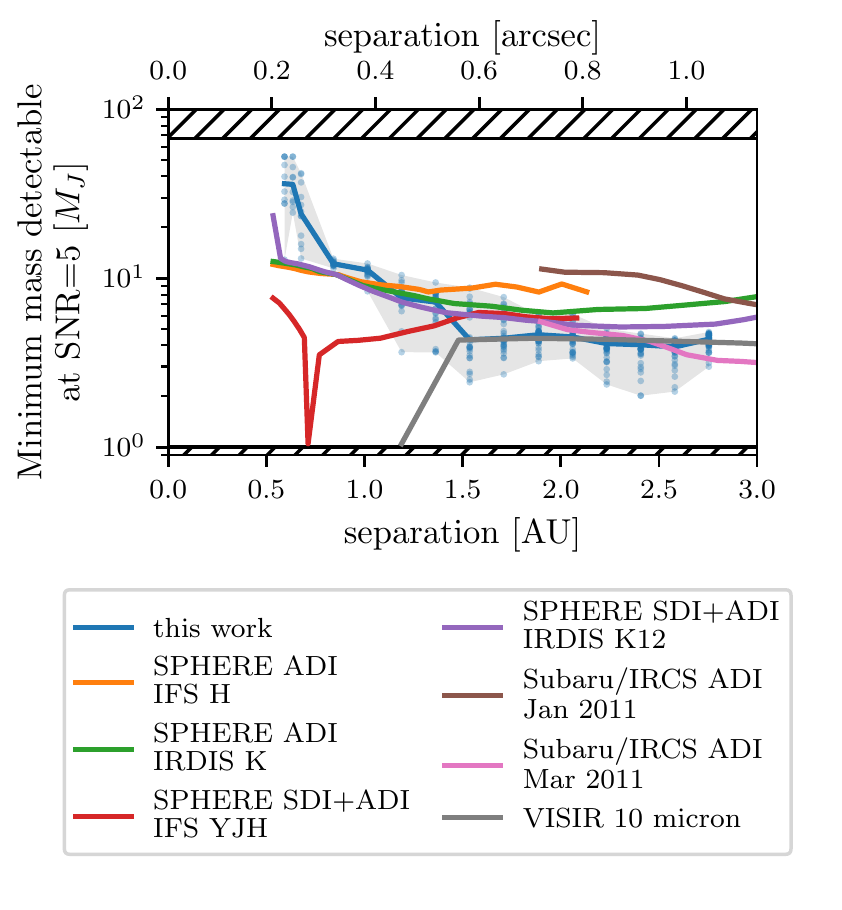}
    \caption{Minimum masses detectable at SNR=5 at our contrast levels using the \cite{marleyBobcat} models when assuming a second epoch of planet formation at $t = 141$ Myr. Blue line and points are from this work. The contrast curves from the literature have also been re-analyzed using the Bobcat models for spectra and evolution, incorporating a $T_\mathrm{eq}$ correction as described in Sec.~\ref{sec:irradiation}. (Points at 1~$M_J$ are likely overestimates, as that is the limit of the model grid.)\label{fig:mass-limit-irr-second-epoch}}
\end{figure}

\begin{figure}
    \includegraphics{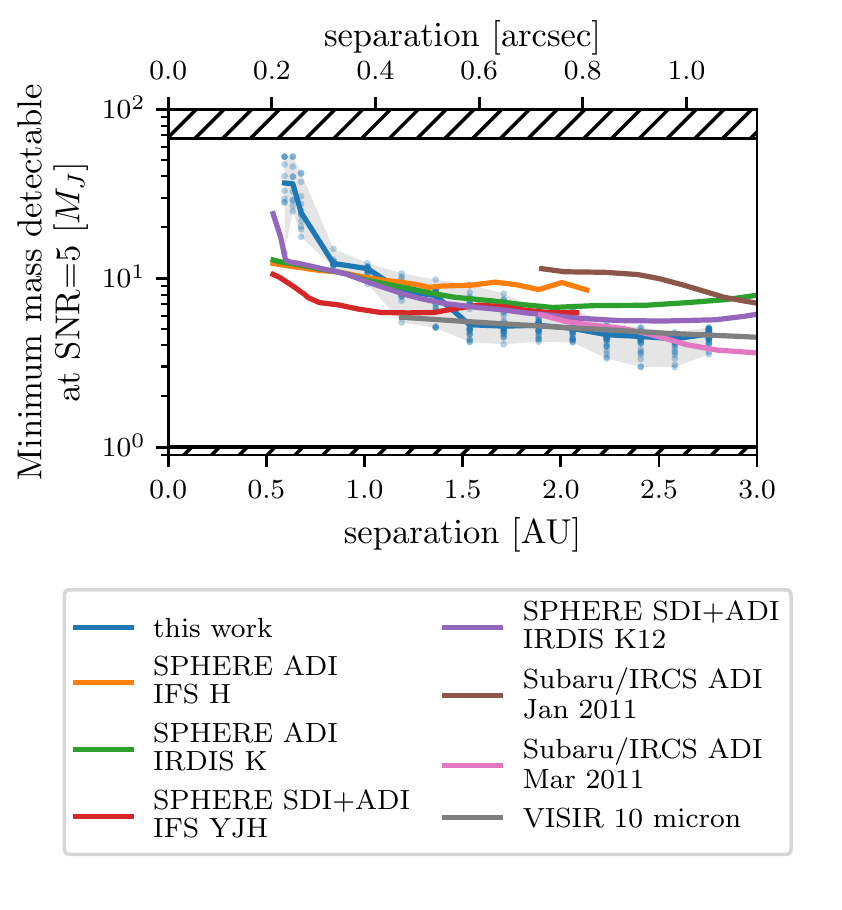}
    \caption{Minimum masses detectable at SNR=5 at our contrast levels using the \cite{marleyBobcat} models assuming a second epoch of planet formation, without irradiation. Blue line and points are from this work.\label{fig:mass-limit-no-irr-second-epoch}}
\end{figure}

The previous analysis assumed that any planets orbiting Sirius A would have formed contemporaneously with their host star. The less time that has elapsed since the time of formation, naturally the brighter a self-luminous giant planet would be, and thus easier for us to detect. If the red-giant stage of Sirius B altered the circumstellar environment of Sirius A to trigger a second epoch of planet formation, would we be sensitive to it?

\begin{figure}
    \includegraphics{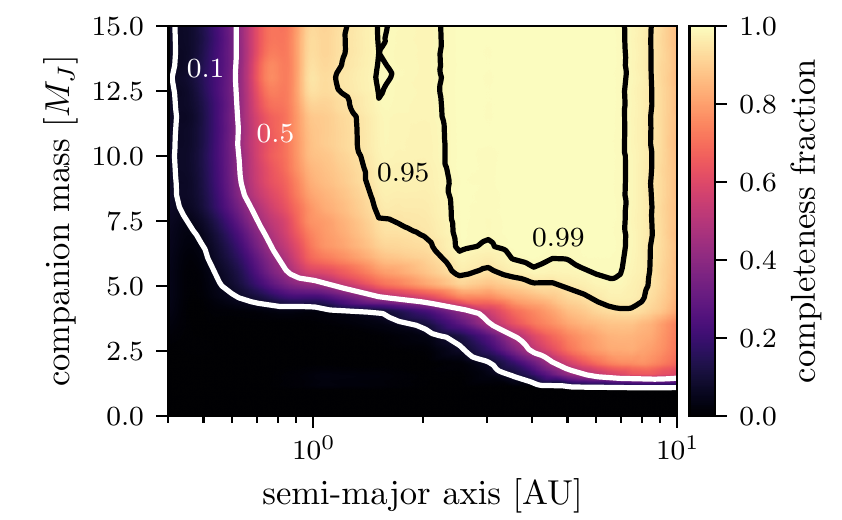}
    \caption{The fraction of planets formed in a hypothetical second period of formation at $t = 141$~Myr on randomly drawn orbits around Sirius A that would be detectable in these observations, as a function of semi-major axis and mass. \label{fig:completeness-141Myr}}
\end{figure}

A 101--126 Myr approximate nuclear lifetime is predicted by models of the Sirius B progenitor \citep{liebertAgeProgenitorMass2005}. Since we have adopted 242 Myr for the system age, the more conservative estimate of a second epoch of planet formation would give us an age of 141~Myr. Calculating the synthetic photometry for model planets of this age gives us the predicted mass limits in Figure~\ref{fig:mass-limit-irr-second-epoch} and Figure~\ref{fig:mass-limit-no-irr-second-epoch}. The completeness as a function of mass and separation increases accordingly, as shown in Figure~\ref{fig:completeness-141Myr}.

\section{Conclusions}

We have demonstrated the use of a gvAPP-180 coronagraph to search for planets, allowing us to probe quite small separations from Sirius A. Expressed in $\lambda / D$, the inner working angle advantages of the gvAPP-180 are even more apparent. We have re-analyzed the reported contrast curves from previous high-contrast imaging searches with a common set of models, and report the overall completeness from all four studies.

\jladd{\cite{Bond2017} report that stable planetary orbits around Sirius A exist up to a maximum period of 2.24 years, citing a numerical stability analysis of planets in binary systems by \cite{holmanLongTermStabilityPlanets1999}. Assuming a circular orbit, this corresponds to a semi-major axis of 2.2 AU. The only regions of Figure~\ref{fig:completeness} with high completeness in our synthesis of previous direct-imaging studies are at larger semi-major axis values than this cutoff. This implies that we have barely begun to probe the region where a companion could exist on a stable orbit.}

\jladd{The albedo range reported in \cite{marleyReflectedSpectraAlbedos1999} of $0.39 < A_B < 0.93$ lets us predict equilibrium temperatures for planets at $r = 2.2$ AU of $217\,\mathrm{K} < T_\mathrm{eff} < 372\,\mathrm{K}$. The low-mass end of the Bobcat models is $0.5\,M_J$, which will have cooled to 191 K by 242 Myr. Thus, 217 K as the pessimistic $T_\mathrm{eq}$ is still greater than the temperature predicted by the evolution model. Applying Equation~\ref{eq:teff} predicts $T_\mathrm{eff} = 244 K$ from the evolutionary and equilibrium temperatures. The model spectrum for a $0.5\,M_J$ self-luminous planet divided by a reference Sirius spectrum predicts that the contrast will be lowest---$6\times10^{-6}$---at 4.6 \um{}, when considering the 1--5 \um{} range of near-IR detectors like Clio and JWST NIRCam. Aside from the inflection in contrast ratio around 4.6 \um{}, the contrast ratio continues falling as wavelength increases, but achieving the required contrast and inner working angle becomes more challenging.}

\jladd{In our data reduction,} we have shown that retaining the full dataset rather than co-adding can result in improved contrast limits. This creates challenges from the increased data volume, but we have shown that the time-domain PCA-based starlight subtraction algorithm PCAT provides advantages in computational cost on large datasets.

The gvAPP-180 coronagraph in Clio, the ancestor of updated designs like those in ERIS and LBT \citep{kenworthyHighContrastImaging2018, doelmanVectorapodizingPhasePlate2021}, presents the same challenges in data reduction common to those coronagraphs: chiefly, dealing with the paired, non-radially-symmetric PSF. We evaluated multiple methods of combining the data, whether performing starlight subtraction on both full PSFs, analyzing each dark-hole region separately, or---as we finally concluded was optimal---dividing out the AO-performance-related variation in the primary star to normalize the dark hole regions into relative units, and then constructing a single observation vector using both dark hole regions.

The vAPP coronagraph makes the limitations of the 1-D contrast limit curve apparent when it comes to capturing the actual sensitivity of a set of observations. Other asymmetric coronagraphs like the PAPLC \citep{porPhaseapodizedpupilLyotCoronagraphs2020,haffertAdvancedWavefrontSensing2022} will also provide similarly complex contrast surfaces, and part of this work's contribution is as an example of quantifying and visualizing that variation.

The appeal of direct imaging studies of Sirius is no doubt in part due to its many convenient properties: brightness, proximity, and a wealth of existing work. However, successive infrared direct-imaging studies have not revealed any new companions, and the contour of 95\% completeness pushes ever inwards and downwards in mass sensitivity. Still, the absence of a detection with current facilities cannot be used to rule out the existence of any planet entirely. For example, a Jupiter-mass planet on a Jupiter-like orbit cannot be ruled out even after synthesizing the results of these previous studies. And, of course, reflected-light exoplanet imaging enabled by next-generation ground-based telescopes will probe a new mass/separation regime altogether.

\section*{acknowledgments}
Support for JRM to conduct these observations was provided, in part, under contract with the California Institute of Technology (Caltech)/Jet Propulsion Laboratory (JPL) funded by NASA through the Sagan Fellowship Program executed by the NASA Exoplanet Science Institute. KMM's and LMC's work was supported in part by the NASA Exoplanets Research Program (XRP) by cooperative agreement NNX16AD44G.

This work has been supported by the Heising-Simons Foundation award \#2020-1824 and NSF MRI Award \#1625441 (MagAO-X). Parts of this research were done using services provided by the OSG Consortium \citep{osg07,osg09}, which is supported by the National Science Foundation awards \#2030508 and \#1836650. S.Y.H. is supported by NASA through the NASA Hubble Fellowship grant \#HST-HF2-51436.001-A awarded by the Space Telescope Science Institute, which is operated by the Association of Universities for Research in Astronomy, Incorporated, under NASA contract NAS5-26555. This work also made use of High Performance Computing (HPC) resources supported by the University of Arizona.

\jladd{The authors would also like to thank the anonymous reviewer for their comments which have improved this manuscript.}

\vspace{5mm}
\facilities{Magellan:Clay (MagAO, Clio)}

\software{
    Astropy \citep{astropy2018},
    Ray \citep{rayPaper},
    POPPY \citep{POPPY},
    matplotlib \citep{matplotlib}
}

\appendix

\section{Spatial distributions of hyperparameters found}\label{sec:hyperparam-distributions}

\begin{figure}
    \gridline{
        \fig{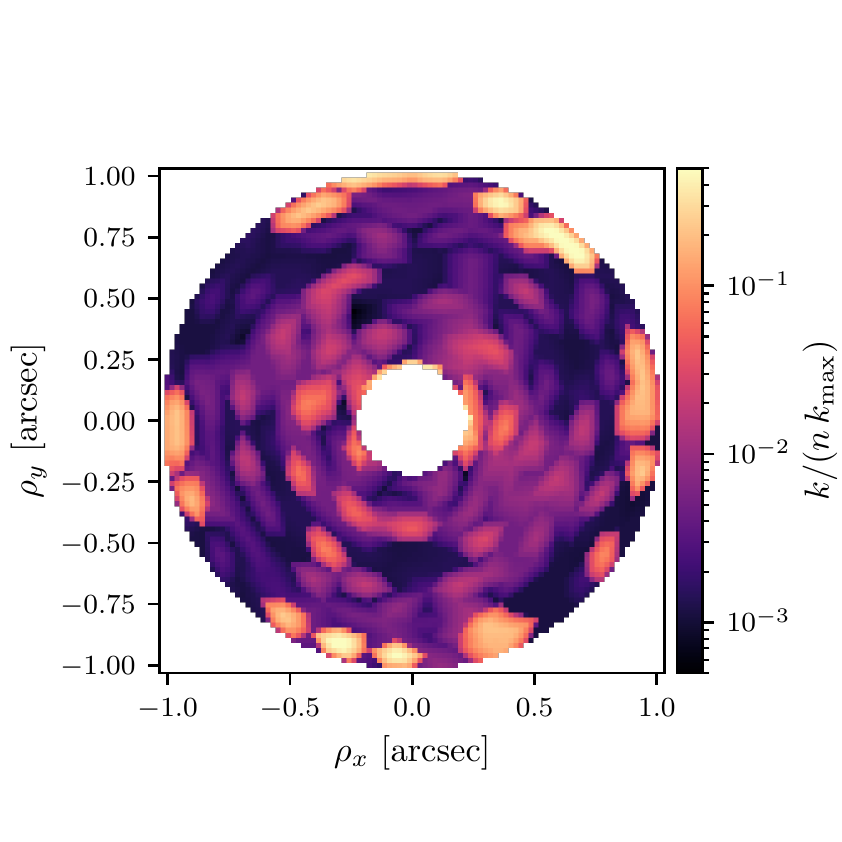}{2.5in}{(a)}
        \fig{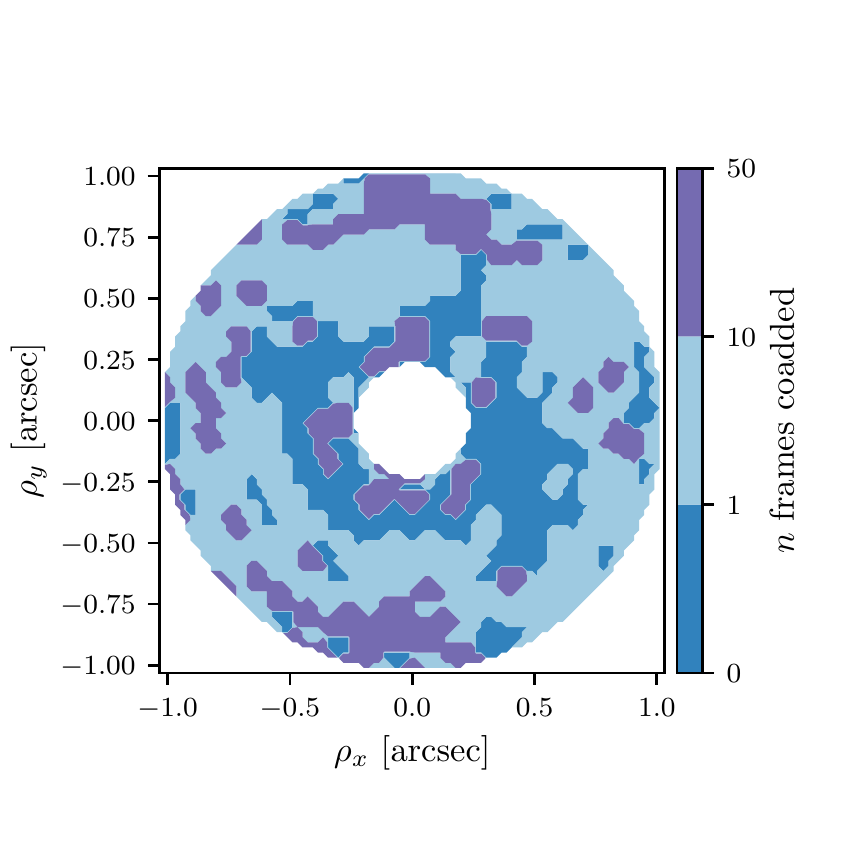}{2.5in}{(b)}
    }
    \gridline{
        \fig{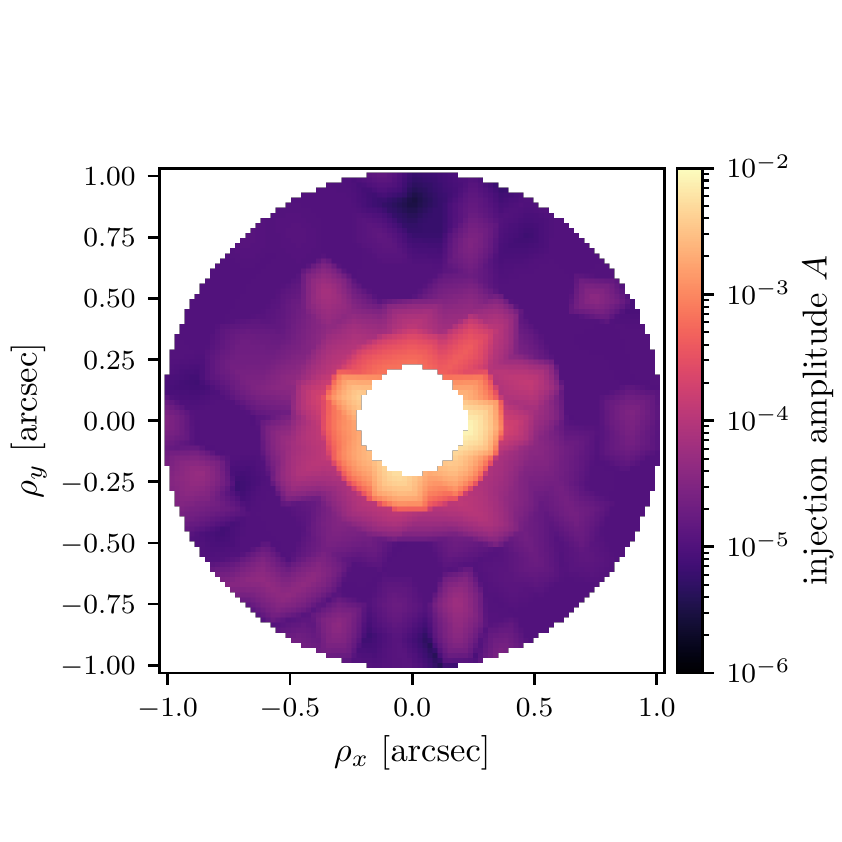}{2.5in}{(c)}
    }
    \caption{\jladd{Visualization of the spatial distributions of hyperparameters found by the optimization process. (a) Distribution of the optimal modes fraction $k/k_\mathrm{max}$ at each position scaled by $1/n$ where $n$ is the optimal number of frames to coadd from the $n \in \{1, 10, 50\}$ values evaluated. (b) Spatial distribution the optimal number of frames to coadd $n \in \{1, 10, 50\}$. (c) Optimized amplitude of injected signal used to measure an SNR and compute the limiting amplitude where SNR=5. Since injecting a signal with an amplitude much greater than the limit will cause us to overestimate the limit, we vary the amplitude in the optimization process as well as described in Section~\ref{sec:hyperparams}.}\label{fig:hyperparam-distributions}}
\end{figure}

\jladd{Figure~\ref{fig:hyperparam-distributions} shows the distribution across the focal plane of the hyperparameters tuned by the optimization process. The maximum number of modes $k_\mathrm{max}$ is different at different locations in the focal plane due to the number of possible modes changing with the number of pixels masked out by the annular mask, as well as the total number of frames changing with coadding. For instance, when the number of frames is limiting the number of modes, a mode fraction of $k/k_\mathrm{max} = 0.5$ could be $\approx 5000$ modes with no coadding or $\approx 100$ modes with 50:1 coadding. Furthermore, the optimizer is not well-suited to integer-valued hyperparameters so it works with $k / k_\mathrm{max}$, which is then converted into a number of modes to perform the starlight subtraction within the function to be optimized. For this reason, Figure~\ref{fig:hyperparam-distributions}(a) shows the mode fraction divided by the number of coadded frames to give a more accurate impression of the number of modes rather than the mode fraction.}

\jladd{The coadding fraction was grid searched rather than smoothly varied, so Figure~\ref{fig:hyperparam-distributions}(b) is shown without any attempt to smoothly interpolate between sample points. It is perhaps noteworthy that there are regions at $\theta = 0$ and $180^\circ$ where the coverage was lowest (see Figure~\ref{fig:coverage}) that the optimizer favored the maximum amount of coadding.}

\bibliography{sources}{}

\begin{thebibliography}{}
\expandafter\ifx\csname natexlab\endcsname\relax\def\natexlab#1{#1}\fi
\providecommand{\url}[1]{\href{#1}{#1}}
\providecommand{\dodoi}[1]{doi:~\href{http://doi.org/#1}{\nolinkurl{#1}}}
\providecommand{\doeprint}[1]{\href{http://ascl.net/#1}{\nolinkurl{http://ascl.net/#1}}}
\providecommand{\doarXiv}[1]{\href{https://arxiv.org/abs/#1}{\nolinkurl{https://arxiv.org/abs/#1}}}

\bibitem[{{Astropy Collaboration} {et~al.}(2018){Astropy Collaboration},
  {Price-Whelan}, {Sip{\H{o}}cz}, {G{\"u}nther}, {Lim}, {Crawford}, {Conseil},
  {Shupe}, {Craig}, {Dencheva}, {Ginsburg}, {VanderPlas}, {Bradley},
  {P{\'e}rez-Su{\'a}rez}, {de Val-Borro}, {Aldcroft}, {Cruz}, {Robitaille},
  {Tollerud}, {Ardelean}, {Babej}, {Bach}, {Bachetti}, {Bakanov}, {Bamford},
  {Barentsen}, {Barmby}, {Baumbach}, {Berry}, {Biscani}, {Boquien}, {Bostroem},
  {Bouma}, {Brammer}, {Bray}, {Breytenbach}, {Buddelmeijer}, {Burke},
  {Calderone}, {Cano Rodr{\'\i}guez}, {Cara}, {Cardoso}, {Cheedella}, {Copin},
  {Corrales}, {Crichton}, {D'Avella}, {Deil}, {Depagne}, {Dietrich}, {Donath},
  {Droettboom}, {Earl}, {Erben}, {Fabbro}, {Ferreira}, {Finethy}, {Fox},
  {Garrison}, {Gibbons}, {Goldstein}, {Gommers}, {Greco}, {Greenfield},
  {Groener}, {Grollier}, {Hagen}, {Hirst}, {Homeier}, {Horton}, {Hosseinzadeh},
  {Hu}, {Hunkeler}, {Ivezi{\'c}}, {Jain}, {Jenness}, {Kanarek}, {Kendrew},
  {Kern}, {Kerzendorf}, {Khvalko}, {King}, {Kirkby}, {Kulkarni}, {Kumar},
  {Lee}, {Lenz}, {Littlefair}, {Ma}, {Macleod}, {Mastropietro}, {McCully},
  {Montagnac}, {Morris}, {Mueller}, {Mumford}, {Muna}, {Murphy}, {Nelson},
  {Nguyen}, {Ninan}, {N{\"o}the}, {Ogaz}, {Oh}, {Parejko}, {Parley}, {Pascual},
  {Patil}, {Patil}, {Plunkett}, {Prochaska}, {Rastogi}, {Reddy Janga},
  {Sabater}, {Sakurikar}, {Seifert}, {Sherbert}, {Sherwood-Taylor}, {Shih},
  {Sick}, {Silbiger}, {Singanamalla}, {Singer}, {Sladen}, {Sooley},
  {Sornarajah}, {Streicher}, {Teuben}, {Thomas}, {Tremblay}, {Turner},
  {Terr{\'o}n}, {van Kerkwijk}, {de la Vega}, {Watkins}, {Weaver}, {Whitmore},
  {Woillez}, {Zabalza}, \& {Astropy Contributors}}]{astropy2018}
{Astropy Collaboration}, {Price-Whelan}, A.~M., {Sip{\H{o}}cz}, B.~M., {et~al.}
  2018, \aj, 156, 123, \dodoi{10.3847/1538-3881/aabc4f}

\bibitem[{{Bailer-Jones} {et~al.}(2021){Bailer-Jones}, Rybizki, Fouesneau,
  Demleitner, \& Andrae}]{gaiaDistances}
{Bailer-Jones}, C. A.~L., Rybizki, J., Fouesneau, M., Demleitner, M., \&
  Andrae, R. 2021, The Astronomical Journal, 161, 147,
  \dodoi{10.3847/1538-3881/abd806}

\bibitem[{Benest \& Duvent(1995)}]{Benest1995}
Benest, D., \& Duvent, J.~L. 1995, Astronomy {\&} Astrophysics, 299, 621

\bibitem[{{Bessel}(1844)}]{Bessel1844}
{Bessel}, F.~W. 1844, \mnras, 6, 136, \dodoi{10.1093/mnras/6.11.136}

\bibitem[{Bohlin {et~al.}(2019)Bohlin, Deustua, \& {de
  Rosa}}]{calspecSiriusUpdate2019}
Bohlin, R.~C., Deustua, S.~E., \& {de Rosa}, G. 2019, The Astronomical Journal,
  158, 211, \dodoi{10.3847/1538-3881/ab480c}

\bibitem[{Bohlin {et~al.}(2014)Bohlin, Gordon, \& Tremblay}]{calspec}
Bohlin, R.~C., Gordon, K.~D., \& Tremblay, P.-E. 2014, Publications of the
  Astronomical Society of the Pacific, 000, \dodoi{10.1086/677655}

\bibitem[{{Bond}(1862)}]{Bond1862}
{Bond}, G. 1862, Astronomische Nachrichten, 57, 131

\bibitem[{{Bond} {et~al.}(2017){Bond}, {Schaefer}, {Gilliland}, {Holberg},
  {Mason}, {Lindenblad}, {Seitz-McLeese}, {Arnett}, {Demarque}, {Spada},
  {Young}, {Barstow}, {Burleigh}, \& {Gudehus}}]{Bond2017}
{Bond}, H.~E., {Schaefer}, G.~H., {Gilliland}, R.~L., {et~al.} 2017, \apj, 840,
  70, \dodoi{10.3847/1538-4357/aa6af8}

\bibitem[{Close {et~al.}(2013)Close, Males, Morzinski, Kopon, Follette,
  Rodigas, Hinz, Wu, Puglisi, Esposito, Riccardi, Pinna, Xompero, Briguglio,
  Uomoto, \& Hare}]{closeDIFFRACTIONLIMITEDVISIBLELIGHT2013}
Close, L.~M., Males, J.~R., Morzinski, K., {et~al.} 2013, The Astrophysical
  Journal, 774, 94, \dodoi{10.1088/0004-637X/774/2/94}

\bibitem[{Doelman {et~al.}(2021)Doelman, Snik, Por, Bos, Otten, Kenworthy,
  Haffert, Wilby, Bohn, Sutlieff, Miller, Ouellet, {de Boer}, Keller, Escuti,
  Shi, Warriner, Hornburg, Birkby, Males, Morzinski, Close, Codona, Long,
  Schatz, Lumbres, Rodack, Van~Gorkom, Hedglen, Guyon, Lozi, Groff, Chilcote,
  Jovanovic, Thibault, {de Jonge}, Allain, Vall{\'e}e, Patel, C{\^o}t{\'e},
  Marois, Hinz, Stone, Skemer, Briesemeister, Boehle, Glauser, Taylor, Baudoz,
  Huby, Absil, Carlomagno, \& Delacroix}]{doelmanVectorapodizingPhasePlate2021}
Doelman, D.~S., Snik, F., Por, E.~H., {et~al.} 2021, Applied Optics, 60, D52,
  \dodoi{10.1364/AO.422155}

\bibitem[{{Ghezzi} {et~al.}(2018){Ghezzi}, {Montet}, \& {Johnson}}]{Ghezzi2018}
{Ghezzi}, L., {Montet}, B.~T., \& {Johnson}, J.~A. 2018, \apj, 860, 109,
  \dodoi{10.3847/1538-4357/aac37c}

\bibitem[{Haffert {et~al.}(2022)Haffert, Males, Van~Gorkom, Close, Long,
  Hedglen, Ahn, Guyon, Schatz, Kautz, Lumbres, Rodack, Knight, Sun, Fogarty, \&
  Miller}]{haffertAdvancedWavefrontSensing2022}
Haffert, S.~Y., Males, J., Van~Gorkom, K., {et~al.} 2022, in Adaptive {{Optics
  Systems VIII}}, ed. D.~Schmidt, L.~Schreiber, \& E.~Vernet ({Montr\'eal,
  Canada}: {SPIE}), 305, \dodoi{10.1117/12.2630425}

\bibitem[{Holman \& Wiegert(1999)}]{holmanLongTermStabilityPlanets1999}
Holman, M.~J., \& Wiegert, P.~A. 1999, The Astronomical Journal, 117, 621,
  \dodoi{10.1086/300695}

\bibitem[{Hunter(2007)}]{matplotlib}
Hunter, J.~D. 2007, Computing in Science \& Engineering, 9, 90,
  \dodoi{10.1109/MCSE.2007.55}

\bibitem[{{Hunziker, S.} {et~al.}(2018){Hunziker, S.}, {Quanz, S. P.}, {Amara,
  A.}, \& {Meyer, M. R.}}]{Hunziker2018}
{Hunziker, S.}, {Quanz, S. P.}, {Amara, A.}, \& {Meyer, M. R.} 2018, A\&A, 611,
  A23, \dodoi{10.1051/0004-6361/201731428}

\bibitem[{{Johnson} {et~al.}(2007){Johnson}, {Fischer}, {Marcy}, {Wright},
  {Driscoll}, {Butler}, {Hekker}, {Reffert}, \& {Vogt}}]{Johnson2007}
{Johnson}, J.~A., {Fischer}, D.~A., {Marcy}, G.~W., {et~al.} 2007, \apj, 665,
  785, \dodoi{10.1086/519677}

\bibitem[{Jones {et~al.}(2013)Jones, Noll, Kausch, Szyszka, \&
  Kimeswenger}]{jonesAdvancedScatteredMoonlight2013}
Jones, A., Noll, S., Kausch, W., Szyszka, C., \& Kimeswenger, S. 2013,
  Astronomy \& Astrophysics, 560, A91, \dodoi{10.1051/0004-6361/201322433}

\bibitem[{{Kenworthy} {et~al.}(2007){Kenworthy}, {Codona}, {Hinz}, {Angel},
  {Heinze}, \& {Sivanandam}}]{Kenworthy2007}
{Kenworthy}, M.~A., {Codona}, J.~L., {Hinz}, P.~M., {et~al.} 2007, \apj, 660,
  762, \dodoi{10.1086/513596}

\bibitem[{Kenworthy {et~al.}(2018)Kenworthy, Snik, Keller, Doelman, Por, Absil,
  Carlomagno, Karlsson, Huby, Glauser, Quanz, \&
  Taylor}]{kenworthyHighContrastImaging2018}
Kenworthy, M.~A., Snik, F., Keller, C.~U., {et~al.} 2018, in Ground-Based and
  {{Airborne Instrumentation}} for {{Astronomy VII}}, ed. H.~Takami, C.~J.
  Evans, \& L.~Simard ({Austin, United States}: {SPIE}), 151,
  \dodoi{10.1117/12.2313964}

\bibitem[{Kervella {et~al.}(2003)Kervella, Th{\'e}venin, Morel, Bord{\'e}, \&
  Di~Folco}]{kervellaSirius}
Kervella, P., Th{\'e}venin, F., Morel, P., Bord{\'e}, P., \& Di~Folco, E. 2003,
  Astronomy \& Astrophysics, 408, 681, \dodoi{10.1051/0004-6361:20030994}

\bibitem[{Krige(1951)}]{krige1951statistical}
Krige, D.~G. 1951, PhD thesis, University of the Witwatersrand

\bibitem[{Li {et~al.}(2018)Li, Jiang, West, Gierasch, {Perez-Hoyos},
  {Sanchez-Lavega}, Fletcher, Fortney, Knowles, Porco, Baines, Fry, Mallama,
  Achterberg, Simon, Nixon, Orton, Dyudina, Ewald, \&
  Schmude}]{liLessAbsorbedSolar2018}
Li, L., Jiang, X., West, R.~A., {et~al.} 2018, Nature Communications, 9, 3709,
  \dodoi{10.1038/s41467-018-06107-2}

\bibitem[{Liebert {et~al.}(2005)Liebert, Young, Arnett, Holberg, \&
  Williams}]{liebertAgeProgenitorMass2005}
Liebert, J., Young, P.~A., Arnett, D., Holberg, J.~B., \& Williams, K.~A. 2005,
  The Astrophysical Journal, 630, L69, \dodoi{10.1086/462419}

\bibitem[{{Long} \& {Males}(2021)}]{accelerateKlip}
{Long}, J.~D., \& {Males}, J.~R. 2021, \aj, 161, 166,
  \dodoi{10.3847/1538-3881/abe12b}

\bibitem[{{Long} {et~al.}(2018){Long}, {Males}, {Morzinski}, {Close}, {Snik},
  {Kenworthy}, {Otten}, {Monnier}, {Tolls}, \& {Weinberger}}]{Long2018}
{Long}, J.~D., {Males}, J.~R., {Morzinski}, K.~M., {et~al.} 2018, in Society of
  Photo-Optical Instrumentation Engineers (SPIE) Conference Series, Vol. 10703,
  \procspie, 107032T, \dodoi{10.1117/12.2312874}

\bibitem[{Males {et~al.}(2021)Males, Fitzgerald, Belikov, \&
  Guyon}]{speckleLives}
Males, J.~R., Fitzgerald, M.~P., Belikov, R., \& Guyon, O. 2021, Publications
  of the Astronomical Society of the Pacific, 133, 104504,
  \dodoi{10.1088/1538-3873/ac0f0c}

\bibitem[{Males {et~al.}(2014)Males, Close, Guyon, Morzinski, Puglisi, Hinz,
  Follette, Monnier, Tolls, Rodigas, Weinberger, Boss, Kopon, Wu, Esposito,
  Riccardi, Xompero, Briguglio, \& Pinna}]{malesDirectImagingExoplanets2014}
Males, J.~R., Close, L.~M., Guyon, O., {et~al.} 2014, in {{SPIE Astronomical
  Telescopes}} + {{Instrumentation}}, ed. E.~Marchetti, L.~M. Close, \& J.-P.
  V{\'e}ran, {Montr\'eal, Quebec, Canada}, 914820, \dodoi{10.1117/12.2057135}

\bibitem[{Marley {et~al.}(1999)Marley, Gelino, Stephens, Lunine, \&
  Freedman}]{marleyReflectedSpectraAlbedos1999}
Marley, M.~S., Gelino, C., Stephens, D., Lunine, J.~I., \& Freedman, R. 1999,
  The Astrophysical Journal, 513, 879, \dodoi{10.1086/306881}

\bibitem[{Marley {et~al.}(2021)Marley, Saumon, Visscher, Lupu, Freedman,
  Morley, Fortney, Seay, Smith, Teal, \& Wang}]{marleyBobcat}
Marley, M.~S., Saumon, D., Visscher, C., {et~al.} 2021, The Astrophysical
  Journal, 920, 85, \dodoi{10.3847/1538-4357/ac141d}

\bibitem[{{Marois} {et~al.}(2006){Marois}, {Lafreni{\`e}re}, {Doyon},
  {Macintosh}, \& {Nadeau}}]{Marois2006}
{Marois}, C., {Lafreni{\`e}re}, D., {Doyon}, R., {Macintosh}, B., \& {Nadeau},
  D. 2006, \apj, 641, 556, \dodoi{10.1086/500401}

\bibitem[{Mawet {et~al.}(2014)Mawet, Milli, Wahhaj, Pelat, Absil, Delacroix,
  Boccaletti, Kasper, Kenworthy, Marois, Mennesson, \& Pueyo}]{mawetSNR}
Mawet, D., Milli, J., Wahhaj, Z., {et~al.} 2014, The Astrophysical Journal,
  792, 97, \dodoi{10.1088/0004-637X/792/2/97}

\bibitem[{Moritz {et~al.}(2018)Moritz, Nishihara, Wang, Tumanov, Liaw, Liang,
  Elibol, Yang, Paul, Jordan, \& Stoica}]{rayPaper}
Moritz, P., Nishihara, R., Wang, S., {et~al.} 2018, in 13th USENIX Symposium on
  Operating Systems Design and Implementation (OSDI 18) (Carlsbad, CA: USENIX
  Association), 561--577.
\newblock \url{https://www.usenix.org/conference/osdi18/presentation/moritz}

\bibitem[{Morzinski {et~al.}(2015)Morzinski, Males, Skemer, Close, Hinz,
  Rodigas, Puglisi, Esposito, Riccardi, Pinna, Xompero, Briguglio, Bailey,
  Follette, Kopon, Weinberger, \& Wu}]{Morzinski2015}
Morzinski, K.~M., Males, J.~R., Skemer, A.~J., {et~al.} 2015, The Astrophysical
  Journal, 815, 108.
\newblock \url{http://stacks.iop.org/0004-637X/815/i=2/a=108}

\bibitem[{Nielsen {et~al.}(2019)Nielsen, De~Rosa, Macintosh, Wang, Ruffio,
  Chiang, Marley, Saumon, Savransky, Mark~Ammons, Bailey, Barman, Blain,
  Bulger, Burrows, Chilcote, Cotten, Czekala, Doyon, Duch{\^e}ne, Esposito,
  Fabrycky, Fitzgerald, Follette, Fortney, Gerard, Goodsell, Graham, Greenbaum,
  Hibon, Hinkley, Hirsch, Hom, Hung, Ilene~Dawson, Ingraham, Kalas, Konopacky,
  Larkin, Lee, Lin, Maire, Marchis, Marois, Metchev, {Millar-Blanchaer},
  Morzinski, Oppenheimer, Palmer, Patience, Perrin, Poyneer, Pueyo, Rafikov,
  Rajan, Rameau, Rantakyr{\"o}, Ren, Schneider, Sivaramakrishnan, Song,
  Soummer, Tallis, Thomas, {Ward-Duong}, \&
  Wolff}]{nielsenGeminiPlanetImager2019}
Nielsen, E.~L., De~Rosa, R.~J., Macintosh, B., {et~al.} 2019, The Astronomical
  Journal, 158, 13, \dodoi{10.3847/1538-3881/ab16e9}

\bibitem[{Noll {et~al.}(2012)Noll, Kausch, Barden, Jones, Szyszka, Kimeswenger,
  \& Vinther}]{nollAtmosphericRadiationModel2012}
Noll, S., Kausch, W., Barden, M., {et~al.} 2012, Astronomy \& Astrophysics,
  543, A92, \dodoi{10.1051/0004-6361/201219040}

\bibitem[{Otten {et~al.}(2017)Otten, Snik, Kenworthy, Keller, Males, Morzinski,
  Close, Codona, Hinz, Hornburg, Brickson, \& Escuti}]{Otten2017}
Otten, G. P. P.~L., Snik, F., Kenworthy, M.~A., {et~al.} 2017, The
  Astrophysical Journal, 834, 175, \dodoi{10.3847/1538-4357/834/2/175}

\bibitem[{Pathak {et~al.}(2021)Pathak, {Petit dit de la Roche}, Kasper,
  Sterzik, Absil, Boehle, Feng, Ivanov, Janson, Jones, Kaufer, K{\"a}ufl,
  Maire, Meyer, Pantin, Siebenmorgen, {van den Ancker}, \&
  Viswanath}]{pathakHighcontrastImagingTen2021}
Pathak, P., {Petit dit de la Roche}, D. J.~M., Kasper, M., {et~al.} 2021,
  Astronomy \& Astrophysics, 652, A121, \dodoi{10.1051/0004-6361/202140529}

\bibitem[{{Perrin} {et~al.}(2016){Perrin}, {Long}, {Douglas},
  {Sivaramakrishnan}, \& {Slocum}}]{POPPY}
{Perrin}, M., {Long}, J., {Douglas}, E., {Sivaramakrishnan}, A., \& {Slocum},
  C. 2016, {POPPY: Physical Optics Propagation in PYthon}, Astrophysics Source
  Code Library.
\newblock \doeprint{1602.018}

\bibitem[{Por(2020)}]{porPhaseapodizedpupilLyotCoronagraphs2020}
Por, E.~H. 2020, The Astrophysical Journal, 888, 127,
  \dodoi{10.3847/1538-4357/ab3857}

\bibitem[{Pordes {et~al.}(2007)Pordes, Petravick, Kramer, Olson, Livny, Roy,
  Avery, Blackburn, Wenaus, W{\"u}rthwein, Foster, Gardner, Wilde, Blatecky,
  McGee, \& Quick}]{osg07}
Pordes, R., Petravick, D., Kramer, B., {et~al.} 2007, in 78, Vol.~78, J. Phys.
  Conf. Ser., 012057, \dodoi{10.1088/1742-6596/78/1/012057}

\bibitem[{Samland {et~al.}(2021)Samland, Bouwman, Hogg, Brandner, Henning, \&
  Janson}]{trap}
Samland, M., Bouwman, J., Hogg, D.~W., {et~al.} 2021, Astronomy \&
  Astrophysics, 646, A24, \dodoi{10.1051/0004-6361/201937308}

\bibitem[{{Schroeder} {et~al.}(2000){Schroeder}, {Golimowski}, {Brukardt},
  {Burrows}, {Caldwell}, {Fastie}, {Ford}, {Hesman}, {Kletskin}, {Krist},
  {Royle}, \& {Zubrowski}}]{Schroeder2000}
{Schroeder}, D.~J., {Golimowski}, D.~A., {Brukardt}, R.~A., {et~al.} 2000, \aj,
  119, 906, \dodoi{10.1086/301227}

\bibitem[{Sfiligoi {et~al.}(2009)Sfiligoi, Bradley, Holzman, Mhashilkar, Padhi,
  \& Wurthwein}]{osg09}
Sfiligoi, I., Bradley, D.~C., Holzman, B., {et~al.} 2009, in 2, Vol.~2, 2009
  WRI World Congress on Computer Science and Information Engineering, 428--432,
  \dodoi{10.1109/CSIE.2009.950}

\bibitem[{Snik {et~al.}(2012)Snik, Otten, Kenworthy, Miskiewicz, Escuti,
  Packham, \& Codona}]{snik2012vector}
Snik, F., Otten, G., Kenworthy, M., {et~al.} 2012, in Modern Technologies in
  Space-and Ground-based Telescopes and Instrumentation II, Vol. 8450, SPIE,
  224--234

\bibitem[{Snoek {et~al.}(2012)Snoek, Larochelle, \&
  Adams}]{snoekPracticalBayesianOptimization2012}
Snoek, J., Larochelle, H., \& Adams, R.~P. 2012, in {{NIPS}}'12, Vol.~2,
  Proceedings of the 25th {{International Conference}} on {{Neural Information
  Processing Systems}} ({Lake Tahoe, Nevada}: {Curran Associates Inc.}),
  2951--2959, \dodoi{10.5555/2999325.2999464}

\bibitem[{Soummer {et~al.}(2012)Soummer, Pueyo, \& Larkin}]{Soummer2012}
Soummer, R., Pueyo, L., \& Larkin, J. 2012, The Astrophysical Journal, 755,
  L28, \dodoi{10.1088/2041-8205/755/2/L28}

\bibitem[{Sutlieff {et~al.}(2021)Sutlieff, Bohn, Birkby, Kenworthy, Morzinski,
  Doelman, Males, Snik, Close, Hinz, {et~al.}}]{sutlieff2021high}
Sutlieff, B.~J., Bohn, A.~J., Birkby, J.~L., {et~al.} 2021, Monthly Notices of
  the Royal Astronomical Society, 506, 3224

\bibitem[{Thalmann {et~al.}(2011)Thalmann, Usuda, Kenworthy, Janson, Mamajek,
  Brandner, Dominik, Goto, Hayano, Henning, Hinz, Minowa, \&
  Tamura}]{Thalmann2011}
Thalmann, C., Usuda, T., Kenworthy, M., {et~al.} 2011, The Astrophysical
  Journal, 732, L34, \dodoi{10.1088/2041-8205/732/2/L34}

\bibitem[{{Vigan} {et~al.}(2015){Vigan}, {Gry}, {Salter}, {Mesa}, {Homeier},
  {Moutou}, \& {Allard}}]{Vigan2015}
{Vigan}, A., {Gry}, C., {Salter}, G., {et~al.} 2015, \mnras, 454, 129,
  \dodoi{10.1093/mnras/stv1928}

\bibitem[{Yang \& Shami(2020)}]{yangHyperparameterOptimizationMachine2020}
Yang, L., \& Shami, A. 2020, Neurocomputing, 415, 295,
  \dodoi{10.1016/j.neucom.2020.07.061}

\end{thebibliography}
\bibliographystyle{aasjournal}

\end{document}